\documentclass[aps,prd,preprint,superscriptaddress,nofootinbib]{revtex4-1}
\usepackage{tensor}
\usepackage{amsmath}
\usepackage{amssymb}
\usepackage{mathtools}
\usepackage{graphicx}
\usepackage{wrapfig}
\usepackage{float}
\usepackage[
	colorlinks=true,
	linkcolor=blue,
	citecolor=blue,
	filecolor=magenta,
	urlcolor=blue
]{hyperref}
\usepackage{xcolor}

\newcommand{\Jacobi}[2]{\text{#1}\!\left(#2\right)}

\newcommand{\be}{\begin{equation}}
\newcommand{\ee}{\end{equation}}
\newcommand{\bea}{\begin{eqnarray}}
\newcommand{\eea}{\end{eqnarray}}
\newcommand{\bes}{\begin{subequations}}
\newcommand{\ees}{\end{subequations}}
\newcommand{\la}{\langle}
\newcommand{\ra}{\rangle}

\begin{document}
\title{Semiclassical predictions regarding a preinflationary era and its effects on the power spectrum}
\author{Paul R. Anderson}
\author{Eric D. Carlson}
\author{Taylor M. Ordines}
\author{Bradley Hicks}
\affiliation{Department of Physics, Wake Forest University, 1834 Wake Forest Road, Winston-Salem, North Carolina 27109, USA}
\date{\today}

\begin{abstract}
 An investigation is undertaken into the properties and effects of a preinflationary era during at least part of which semiclassical gravity was valid.	
It is argued that if the Universe (or our part of it) was approximately homogeneous and isotropic during that era, then the Universe was likely to have been radiation dominated.  A simple model in which the Universe contains classical	radiation and a cosmological constant is used to investigate potential effects of such a preinflationary era on the cosmic microwave background.  The power spectrum is computed using the mode functions of a quantized massless minimally coupled scalar field. Various choices of state for this field are considered, including adiabatic vacuum states of various
orders and the vacuum state that would naturally occur if the Universe made a sudden transition from being radiation dominated to de Sitter space.  In all cases investigated,
there is a suppression of the power spectrum at large angles, and, when plotted as a function of the momentum parameter, there are always oscillations with state-dependent amplitudes.
\end{abstract}

\maketitle

\section{Introduction}\label{sec:Introduction}
In the traditional big bang theory, the Universe began with zero size and an initial curvature singularity.
Of course, what this really means is that classical general relativity breaks down, and a description of the very early Universe must come from a quantum theory of gravity.
Since it is currently unknown which, if any, of the current quantum gravity candidates is correct, the beginning of the Universe (if there was one) is unknown.
However, the paradigm today is that early in its history the Universe underwent a period of inflation.
 If there was a preinflationary era, then semiclassical gravity may well have been valid during the latter part of that era.
 One expects that semiclassical gravity would be valid in the early Universe once the curvature is well below the Planck scale.  This would be true for spacetime curvatures of the order of $10^{-4}$ in Planck units which in the early Universe would correspond to energy scales about 100 times larger than that of the grand unified theory, GUT, energy scale.  For this reason it is interesting to explore the predictions that semiclassical gravity makes about the Universe  prior to inflation.

There is some ambiguity as to the exact form of the semiclassical Einstein equations due to the unknown sizes of the coefficients of the scalar curvature squared and Ricci squared terms in the gravitational Lagrangian.
Renormalization of the stress-energy tensor for quantum fields in curved space requires the existence of such terms.
Nevertheless, when gravity is thought of as an effective field theory, one expects that the contributions from such terms to the semiclassical Einstein equations should be relatively small.
If that is the case, and if the preinflationary Universe, or at least our part of it, was approximately homogeneous and isotropic, then from the point of view of semiclassical gravity, the Universe began with zero size as in the classical big bang model.
Of course one of the advantages of inflation is that the part of the Universe we can observe today would have been an extremely small part of the Universe at the onset of inflation.  Thus, if there were significant inhomogeneities on larger scales, they would be well outside the current horizon.
However, here we make the stronger assumption that any large inhomogeneities were far enough away that the part of the Universe that contained the part we can see today, and was significantly larger than it, was approximately homogeneous and isotropic.  Then our argument implies that this portion was radiation dominated at least during the latter part of the preinflationary era.
Given our ignorance of the preinflationary
era, it is of interest to consider models of this type.

It was argued in~\cite{anderson_effects_1985} that if the Universe began with zero size, then it is possible to define an initial vacuum state for a conformally coupled massive scalar field that, at the initial time, is equivalent to the conformal vacuum state for a conformally coupled massless scalar field. This was done by replacing the mode equation with a Volterra equation that could be solved iteratively.  At lowest order it was shown that for any other homogeneous and isotropic state, the stress-energy tensor $\la T_{\mu\nu} \ra$ contains terms that have the same form as classical radiation.  This is not surprising since it is known~\cite{Wald1978} that for a conformally invariant field in any other homogeneous and isotropic state than the conformal vacuum the stress-energy tensor has such terms.

In this paper we investigate the properties of this initial vacuum state as well as other homogeneous and isotropic vacuum states in significantly more detail than was done in~\cite{anderson_effects_1985}.  We show that for the vast majority of cases where the Universe begins with zero size in any other homogeneous and isotropic state, the stress-energy tensor for a massive conformally coupled scalar field at early times has a term that acts like classical radiation.   We use this to make an argument that it is extremely likely that if the Universe had a period before inflation in which the semiclassical approximation was valid, then it expanded in approximately the same way as a radiation-dominated universe during that period.  Evidence for a radiation-dominated preinflationary era has also been
found in a model in which the Wheeler-DeWitt equation is solved in the
minisuperspace approximation which includes the Hamiltonian for the
scale factor when a cosmological constant is present along with the
Hamiltonian for a single mode of a massless minimally coupled scalar
field~\cite{Appignani_2008}.

From an observational point of view, the best chance for evidence of a preinflationary radiation-dominated phase for the Universe would likely come from the cosmic microwave
background, where it has been shown~\cite{anderson_short_2005} that if inflation did not go on for too long, then there could be significant deviations from the usual prediction if the state of the quantum field differs significantly from the Bunch-Davies state~\cite{Nacht,CherTag,Tag,Bunch-Davies}.

With this as motivation we consider a simple model in which the Universe contains classical radiation and a cosmological constant.  At early times the Universe expands like a radiation-dominated universe and at late times like a de Sitter universe.  This model gives a natural onset to inflation described entirely by the cosmological constant.
 Since we are concerned with the effects on the power spectrum of the preinflationary phase, our results are independent of the reheating phase which occurs in most inflationary models and they are also compatible with warm inflation models~\cite{warm-inflation} in which there is a gradual transfer of energy from the inflaton field to the radiation which eliminates the need for a reheating phase.

In many models of inflation the inflaton field is treated as a classical minimally coupled scalar field with a potential while quantum fluctuations of the inflaton field are treated as a quantized massless minimally coupled scalar field.  In this paper we are effectively modeling the classical inflaton field with a cosmological constant.  We compute
the power spectrum using the mode functions for a massless minimally coupled scalar field.
The effects of certain types of initial vacuum states for this field on the power spectrum are investigated.
One is the natural vacuum state that occurs in a pure radiation-dominated universe that suddenly transforms into de Sitter space.  The others
are adiabatic vacuum states~\cite{parker-thesis,pf,fp,fph,bunch} of zeroth, second, and fourth order.
There have been several previous calculations of the power spectrum for various models in which the preinflationary era was homogeneous, isotropic, and
radiation dominated~\cite{hirai_initial_2003,cline_does_2003,hirai_initial_2005,hirai_effect_2006,powell_pre-inflationary-2007,hirai_effect_2007,wang_effects_2008,Nicholson2008_The,marozzi_on_2011,cicoli_just_2014,das_revisiting_2015,Das2016_Constraints}. As is discussed in Sec.~\ref{sec:Results}, it appears that in most previous cases a sudden approximation or something similar to one was used. Two exceptions are Refs.~\cite{wang_effects_2008,Nicholson2008_The}, where the power spectrum was computed numerically using zeroth-order adiabatic states.  A detailed comparison of our results with theirs is given in Sec.~\ref{sec:Results}.

In agreement with previous calculations we find that the power spectrum deviates from that of the Bunch-Davies state because the initial vacuum state differs from the Bunch-Davies state.  In particular the power spectrum is suppressed at large angles.  When plotted in terms of the momentum parameter $k$ there are oscillations for all
of the states considered. The largest oscillations come from the sudden approximation and from zeroth-order adiabatic states where the adiabatic matching time (discussed in Sec.~\ref{sec:Adiabatic}) occurs near the onset of inflation.  For adiabatic states the oscillations have significantly smaller amplitudes for earlier matching times and for higher-order adiabatic states.

In Sec.~\ref{sec:Prediction} we present our argument that if there was a preinflationary phase in which the semiclassical approximation was valid and if the Universe or our part of it was approximately homogeneous and isotropic during that phase, then it is likely that it expanded like a radiation-dominated universe.
In Sec.~\ref{sec:OurModel} we discuss the solution to Einstein's equations for our specific model, which consists of classical radiation and a cosmological constant.
The different states that we use for the computations of the power spectrum of the massless minimally coupled scalar field are discussed in Sec.~\ref{sec:States}.
A general form for the power spectrum for our model is derived in Sec.~\ref{sec:PowerSpectrum}.  Some of our computations of the power spectrum are presented, discussed, and compared with previous calculations in Sec.~\ref{sec:Results}.
A brief summary of our results is given in Sec.~\ref{sec:Conclusions}.
 The Appendix contains details of the calculations related to a possible radiation-dominated preinflationary phase.
Throughout we use units such that $\hbar = c = G = 1$.

\section{Prediction regarding a preinflationary era}\label{sec:Prediction}

As discussed in the introduction, we consider the possibility that the Universe, or our part of it, began with zero size in a homogeneous and isotropic state from the point of view of semiclassical gravity.  We further assume that there was a preinflationary era in which the semiclassical approximation was valid and that during this era interactions between the quantum fields present did not make the dominant contributions to the stress-energy tensors of those fields.  In this case we present an argument that it is very likely the Universe was expanding like a radiation-dominated universe during this preinflationary era.

As mentioned in the introduction, in many models of inflation the inflaton field is a massive minimally coupled scalar field that is treated classically.  Quantum fluctuations of this field during the period of inflation are generally approximated by a quantized massless minimally coupled scalar field.  In such models, the power spectrum is computed from these fluctuations.  This is also our approach here.  However, most quantum fields that are likely to have had a significant impact on the expansion of the Universe in a semiclassical preinflationary era are of spin $\frac{1}{2}$ and spin $1$.  In the approximation that interactions are neglected, the massless ones are exactly conformally invariant and the massive ones are conformally invariant in the limit that their masses go to zero.  Thus the effects of the inflaton field on the expansion during a preinflationary phase are expected to be small.

For conformally invariant fields in a spatially flat homogeneous and isotropic spacetime the stress-energy tensor $\la 0| T_{\mu\nu} |0 \ra$ is composed of two local tensors that contain higher derivative terms~\cite{birrell_quantum_1982}.  If the semiclassical approximation is valid, then it is usually assumed that these terms are very small\footnote{An important exception is Starobinsky inflation~\cite{starobinsky_new_1980}, which requires that the coefficient of the $R^2$ term in the gravitational Lagrangian be of the order of $10^9$ and that it have a certain sign.  We do not consider Starobinsky inflation in this paper.}.  If a conformally invariant field in such a spacetime is in a homogeneous and isotropic state other than the conformal vacuum state, then there is an additional term in its stress-energy tensor that has the same form as that of classical radiation~\cite{Wald1978}.  Therefore, if the early Universe consisted only of massless conformally invariant quantum fields in homogeneous and isotropic states, and if one or more of the fields was not in the conformal vacuum state, then the Universe would expand like a radiation-dominated universe provided the higher derivative terms made a small contribution to the stress-energy tensor.

Of course many of the quantum fields in the early Universe were massive.  We model the spin-$\frac{1}{2}$ and spin-$1$ massive fields with conformally coupled massive scalar
fields. The rationale for doing this is that all of these fields are conformally invariant in the massless limit, and at high enough momenta they are effectively massless.

We also restrict our attention to cases when the Universe begins with zero scale factor.  We do this, as mentioned in the introduction, in the same spirit that classical
general relativity predicts that the Universe began with an initial singularity.

The metric for a spatially flat homogeneous and isotropic universe is
\begin{equation}\label{Metric}
ds^2 = a^2(\eta) \, (-d \eta^2 + d \vec{x}^2) \;,
\end{equation}
with $\eta$ the conformal time defined by $a\,d \eta = dt$.
Scalar fields with arbitrary masses and curvature couplings $\xi$ satisfy the equation
\begin{equation}\label{FieldEquation}
\Box \phi - m^2 \phi - \xi R \phi = 0 \;,
\end{equation}
where the scalar curvature is
\begin{equation}\label{eq:ScalarCurvature}
R = \frac{6 a''}{a^3} \;.
\end{equation}
Here primes denote derivatives with respect to $\eta$.
Expanding the fields in terms of modes in the usual way gives
\begin{equation}
	\phi = \int d^3 k \; \left[a_{\vec{k}} e^{i \vec{k}\cdot \vec{x}} \phi_k(\eta) + a^\dagger_{\vec{k}} e^{-i \vec{k}\cdot \vec{x}} \phi^{*}_k(\eta)\right] \;. 	
\label{PhiModes}
\end{equation}
With the definitions
\begin{subequations}
	\begin{align}
	\begin{split}
		\phi_k &= \frac{\psi_k}{a}\label{PsiDef} \; ,
	\end{split}\\
	\begin{split}
		\omega_k^2 &= k^2 + m^2 a^2  \; ,\label{OmegaDef}
	\end{split}
	\end{align}
\end{subequations}
one finds that $\psi_k$ is a solution to the equation~\cite{birrell_quantum_1982}
\begin{equation}\label{ModeEquation}
		 \psi_k'' + \left[\omega_k^2 + 6 \left(\xi - \frac{1}{6}\right) \frac{a''}{a} \right] \psi_k = 0
\end{equation}
and satisfies the Wronskian condition
	\be \label{Wronskian} 	 \psi_k \psi_k^{*\prime} - \psi_k^{*} \psi_k^\prime = i \;.   \ee

The classical expression for the stress-energy tensor of an arbitrarily coupled scalar field is~\cite{birrell_quantum_1982}
	\bea
		T_{\mu \nu}  &=& (1-2\xi) \partial_\mu \phi \partial_\nu \phi + \left(2 \xi - \frac{1}{2} \right) g_{\mu \nu}
			\left( g^{\rho \sigma} \partial_\rho \phi \partial_\sigma \phi + m^2 \phi^2\right)
			- 2 \xi \phi \nabla_\mu \nabla_\nu \phi  \nonumber \\
		& & \qquad {}+ 2  g_{\mu\nu}\xi^2 R \phi^2+ \xi G_{\mu \nu} \phi^2  \;.  \label{TmnClassical}
	\eea
Substituting \eqref{PhiModes} and \eqref{PsiDef} into \eqref{TmnClassical} then yields an unrenormalized energy density \cite{bunch}:
	\be
		\rho_u = \frac{1}{4\pi^2 a^4} \int dk \, k^2 \biggl\{ \left| \psi_k^\prime\right|^2 + \omega_k^2 \left| \psi_k\right|^2
			+6\left(\xi-\frac16\right) \left[ \frac{a^\prime}{a} \left( \psi_k^\prime \psi_k^* +\psi_k\psi_k^{*\prime} \right)
			- \frac{a^{\prime 2}}{a^2} \left|\psi_k \right|^2 \right]   \biggr\} \; .\label{RhoPsi}
	\ee

Specializing to the case of conformal coupling, $\xi = \frac{1}{6}$, it is helpful to define functions $\alpha_k(\eta)$ and $\beta_k(\eta)$ by the simultaneous equations
\bes\bea
	\psi_k(\eta) &=& \frac{1}{\sqrt{2\omega_k(\eta)}} \left[\alpha_k(\eta) e^{-i\theta_k(\eta)} + \beta_k(\eta) e^{i\theta_k(\eta)} \right] \label{AlphaBetaDef1}\; , \\
	\psi_k^\prime(\eta)  &=& \sqrt{\frac{\omega_k(\eta)}{2}} \left[-i\alpha_k(\eta) e^{-i\theta_k(\eta)} + i\beta_k(\eta) e^{i\theta_k(\eta)} \right] \label{AlphaBetaDef2}
	\; ,
	\eea\ees
where
\be \theta_k(\eta) = \int^\eta dx \, \omega_k(x) \; . \label{theta-def} \ee
The lower limit for this integral is arbitrary.
The Wronskian condition \eqref{Wronskian} becomes
\be \left|\alpha_k\right|^2 - \left|\beta_k\right|^2 = 1 \; . \label{WronskianAlphaBeta} \ee
Substituting \eqref{AlphaBetaDef1} and \eqref{AlphaBetaDef2} into \eqref{RhoPsi}, setting $\xi = \frac{1}{6}$ and using \eqref{WronskianAlphaBeta} yields
\bea
\rho_u &=& \frac{1}{4\pi^2 a^4} \int_0^\infty dk \, k^2 \omega_k\left(1 + 2\left| \beta_k \right|^2 \right) \; .
\label{RhoAlphaBeta}
\eea
Subtracting off the adiabatic counterterms~\cite{bunch,Anderson_1999,Anderson1987_Adiabatic} one finds
\bea \rho_r &=& \frac{1}{2\pi^2 a^4} \int_0^\infty dk\, k^2 \omega_k\,\left| \beta_k \right|^2  - \frac{m^2}{96 \pi^2} \frac{a^{\prime \,2}}{a^4} + \frac{1}{2880 \pi^2} \left( - \frac{1}{6} {}^{(1)}\!{H_0}^{0} + {}^{(3)}\!{H_0}^0 \right)\;.
\label{RhoAlphaBeta} \eea
 The second term on the right is a finite renormalization of ${G_0}^0$, and the last two terms are the higher derivative terms that are assumed to be small.

The first term on the right in~\eqref{RhoAlphaBeta} has the same form as the energy density for classical radiation if $\omega_k|\beta_k|^2$ is independent of time.  However, it is actually a function of time, so its behavior at early times needs to be analyzed.  This is done in the Appendix, where it is shown that if $\beta_k(\eta)$ is nonzero in the limit $a(\eta)\to 0$,
then the initial behavior of the first term in~\eqref{RhoAlphaBeta} is that of classical radiation provided that (i) the integral in \eqref{RhoAlphaBeta} is finite at $\eta_0$;  (ii) $| \beta_k(\eta_0)|$ increases slower than $k^{-1}$ at small $k$; (iii) the derivative $(a^2)^\prime$ has a finite limit as $\eta \rightarrow \eta_0$; and (iv) $\int_{\eta_0}^\eta |(a^2(x))''| dx$ is finite.  If these conditions are satisfied then the resulting solution to the semiclassical Einstein equations will  describe a universe that expands like a radiation-dominated universe at early times since the second term in~\eqref{RhoAlphaBeta} is a finite renormalization of ${G_0}^0$ and the last terms are assumed to be negligible.

\subsection{A ``natural'' vacuum state}
\label{sec:natvacstate}

It is well known that in a dynamical spacetime there is usually no state that one can unambiguously label as the vacuum state as there is for free free quantum fields
in Minkowski space.  However, there can be states which for one reason or another are preferred.  One example is the Bunch-Davies state in pure de Sitter space~\cite{Nacht,CherTag,Tag,Bunch-Davies}.  Another is the class of states found in~\cite{Agullo-Nelson-Ashtekar} for which at a given moment of time the stress-energy tensor for the quantum
field is exactly equal to zero.  Here we discuss a different choice for a vacuum state based on the above analysis of states for a massive conformally coupled scalar field.

The state when $\beta_k(\eta_0) = 0$ for the conformally coupled massive scalar field provides a natural definition of a vacuum state if the Universe began with zero size since, as shown above, there is no term in the energy density that acts like classical radiation.
 One might guess that a similar state would exist, at least in some cases, for nonconformally coupled scalar fields.  This is correct, but as we next show, in some important cases the state is problematic for nonconformally coupled scalar fields and potentially problematic for conformally coupled massive scalar fields.

Returning to \eqref{ModeEquation}, it is useful to define an effective mass
\be\label{MaDef} M_a^2 = m^2 a^2 + 6\left(\xi - \frac16 \right) \frac{a''}{a} \; . \ee
If initially
\be\label{ConformalVacuum} \psi_k = \frac{e^{-i k \eta}}{\sqrt{2 k}} \;,  \ee
which is the exact solution for the conformally invariant scalar field in the conformal vacuum state, then one can find a formal solution in terms of a Volterra equation:
\be\label{Volterra}
\psi_k(\eta) = \frac{1}{\sqrt{2k}} e^{-ik\,\eta} - \frac{1}{k} \int_{\eta_0}^\eta dx_1 M_a^2(x_1) \, \sin\left[k(\eta - x_1)\right]
\psi_k(x_1) \; .
\ee
This can be solved by iteration to give
\bes\bea
	\psi_k(\eta) &=& \frac{e^{-ik\eta}}{\sqrt{2k}} + \sum_{n=1}^\infty \frac{(-1)^n}{k^n} I_n(k,\eta) \; ,
	\label{VolterraSeries1} \\
	I_n(k,\eta) &=& \int_{\eta_0}^\eta dx_1 \int_{\eta_0}^{x_1} dx_2 \cdots \int_{\eta_0}^{x_{n-1}} dx_n M_a^2(x_1)
	\sin\left[k(\eta - x_1)\right] \nonumber \\
	&&\qquad\qquad \times \left\{\prod_{j=2}^n M_a^2(x_j) \sin\left[k(x_{j-1} - x_j)\right] \right\}
	\frac{e^{-ikx_n}}{\sqrt{2k}} \; ,  \label{VolterraSeries2}
	\eea\label{VolterraSeries}\ees
where the product is equal to $1$ for $n = 1$, $a(\eta_0) = 0$, and either $\eta_0 =  -\infty$ or $-\infty < \eta_0 < \infty$.
It can be shown that this converges provided that $k^{-1} \int_{\eta_0}^{\eta} dx |M^2_a(x)|$ is finite~\cite{anderson_effects_1985}.
We shall restrict our attention to those cases in what follows.

The first term in the sum in~\eqref{VolterraSeries1}  can be written as
	\be
		-\frac{1}{k} I_1(k,\eta) =  -\frac{1}{2ik} \frac{e^{i k \eta}}{\sqrt{2k}} \int_{\eta_0}^\eta dx_1 e^{-2 i k x_1}
			M^2_a(x_1) + \frac{1}{2ik} \frac{e^{-i k \eta}}{\sqrt{2k}} \int_{\eta_0}^\eta dx_1
			M^2_a(x_1) \;. \label{I1}
	\ee
The second term on the right is positive frequency for all times.  The first term in some cases has a negative frequency component.  This was not noticed in~\cite{anderson_effects_1985}.
To see this, assume that there is no divergence in $M_a^2$ or any of its derivatives at $\eta_0$.  Then successive integrations by parts can be done.  The evaluation
of each at the upper limit yields a positive frequency term.  The evaluation at the lower limit yields a negative frequency term.  Thus if $M_a^2$ or any of its derivatives
is nonzero at $\eta_0$ then there is a negative frequency term.  Suppose that $M_a^2$ and its first $(n-1)$ derivatives are zero at $\eta_0$ and that  the $n$'th derivative
of $M_a^2$ is nonzero at $\eta_0$.  Then the vacuum state can be of adiabatic order $n-1$ only if $m \ne 0$ and $\xi = \frac{1}{6}$.  The vacuum state can be of adiabatic
order $n+1$ only if $m = 0$ and $\xi \ne \frac{1}{6}$.  If $m \ne 0$ and $\xi \ne \frac{1}{6}$ then the order of the vacuum state depends on whether it is the $n$'th derivative of the $m^2 a^2$ term (giving $n-1$) or the other term in~\eqref{MaDef} (giving $n+1$).

From the point of view of pure mathematics, one can think of Eq.~\eqref{ModeEquation} as a mode equation in flat space with a time-dependent potential.  If the potential vanishes in the limit $\eta_0 = -\infty$, then it turns on very slowly and this leads to an infinite-order initial vacuum state. In the case that $\eta_0$ is finite, the potential turns on at the time $\eta_0$.  How rapidly it turns on depends on how rapidly $a \to 0$.  The more rapidly it turns on, the more particle production one would expect to occur due to the ``turn on'' and the lower the order of the adiabatic state that the vacuum state corresponds to.
Of course there are cases where the potential is a constant in the limit $\eta \to - \infty$ and cases where it (or one of its derivatives) diverges at $\eta = \eta_0 > -\infty$.
We do not consider these cases here.

An important example where the spacetime begins at $\eta_0 = -\infty$ and the vacuum is an infinite-order adiabatic state is de Sitter space in spatially flat coordinates, where $a = \frac{1}{-H \eta}$ with $H$ a constant.  The vacuum state in this case is the Bunch-Davies state.

An important example where the vacuum state is a finite-order adiabatic state is when the scale factor can be expanded in the power series $a(\eta) = \sum_{n=1}^\infty a_n (\eta-\eta_0)^n $ .  In general there are two contributions to $M_a^2$.  One comes from the $m^2 a^2$ term, which occurs for any massive field.  For it, one finds that if $a_1 \ne 0$, so that the
Universe is approximately radiation dominated at early times, then the vacuum state is at most a first-order adiabatic state.  The second contribution to $M_a^2$ is proportional to $\frac{a''}{a}$.  If $a_1 \ne 0$ then the Volterra solution~\eqref{VolterraSeries} does not work unless $a_2 = a_3 = 0$.  In that case, if $m = 0$ and $\xi \ne \frac{1}{6}$ the vacuum state is at most second-order adiabatic if $a_4 \ne 0$, third-order adiabatic if $a_4 =0$ and $a_5 \ne 0$ and so forth.  For the model described in the next section $a_4 = 0$ and $a_5 \ne 0$, so the vacuum state is at most a third-order adiabatic one.

In general it is necessary to have a fourth-order adiabatic state for the stress-energy tensor to be ultraviolet finite.  Thus, at least for the model we consider below, the vacuum state discussed here is not acceptable for the massless minimally coupled scalar field.  For a conformally coupled massive scalar field it is technically necessary only to have a zeroth-order adiabatic state, so this vacuum state could work.  However, if the spacetime is even slightly inhomogeneous or anisotropic, then something akin to a fourth-order adiabatic state would be required to yield a finite stress-energy tensor.  Therefore we do not consider this to be an acceptable vacuum state for a massive field for the model considered below or for any model of the Universe in which the expansion approaches that of a radiation-dominated universe at early times but is not exactly equal to that of a radiation-dominated universe.

\section{Simple model with a radiation-dominated preinflationary era}\label{sec:OurModel}
 For the rest of this paper we consider a simple model that has a radiation-dominated preinflationary era and a late time inflationary era.  It  consists of classical radiation plus a positive cosmological constant $\Lambda$.
 In this case,	one of the Friedmann-Lema\^itre-Robertson-Walker equations is
	\begin{equation}\label{eq:FriedmannOriginal}
		\frac{a^{\prime 2}}{a^4} = \frac{\Lambda}{3} + \frac{8\pi c_r}{3 a^4} \;,
	\end{equation}
	where $c_r > 0$ is a constant.
The trace of the Einstein equations gives
\be R = \frac{6 a''}{a^3} = 4 \Lambda\;.  \label{R-Lambda} \ee

\begin{figure}[h]
	\includegraphics[width=\linewidth]{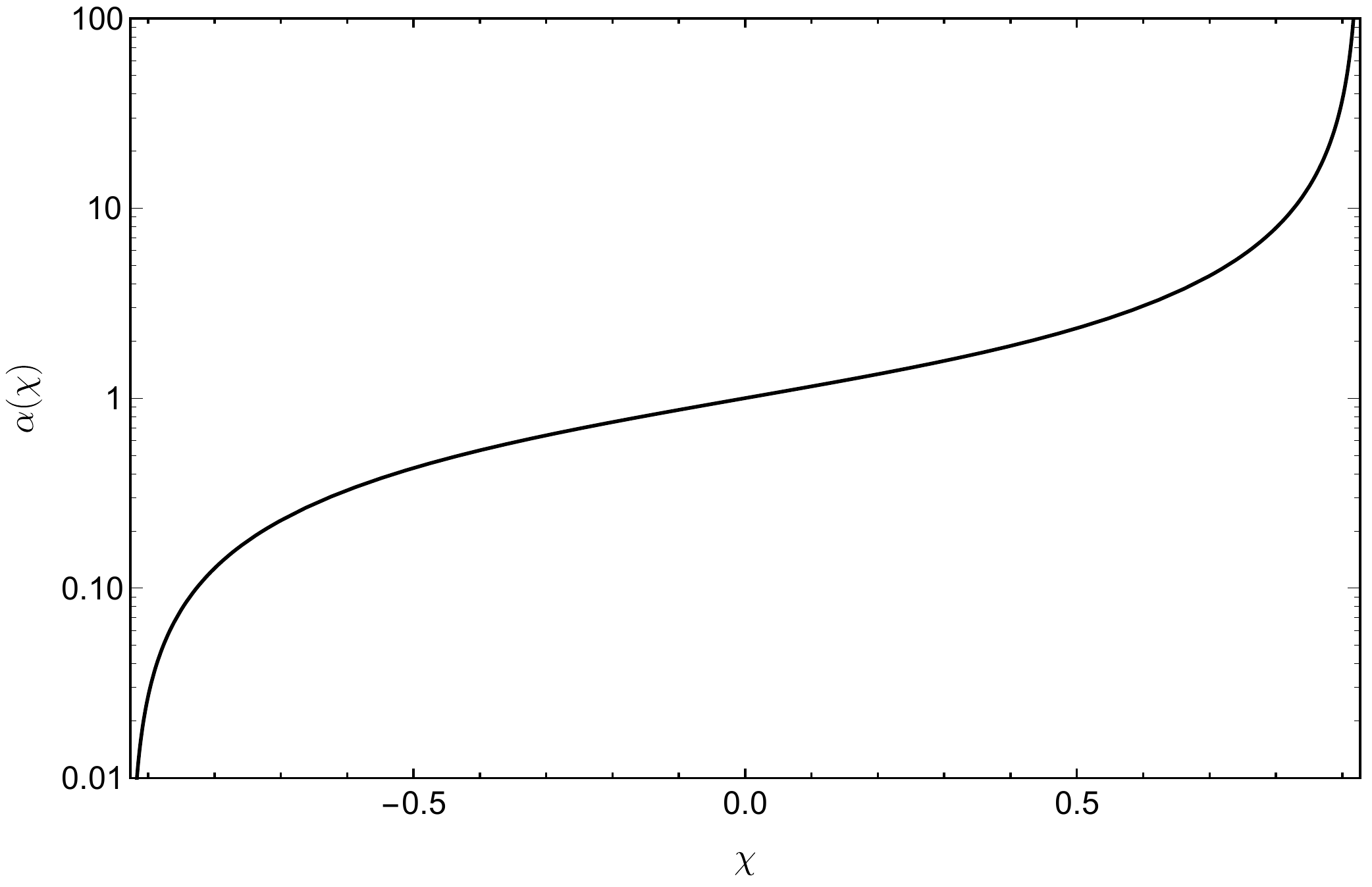}
	\caption{Rescaled scale factor $\alpha(\chi)$ over its domain $(\chi_0,\chi_\infty)$.\label{fig:AlphaPlot}}
\end{figure}

	We use the following scaled variables:
	\bes\label{eq:ScaledVariables}
		\bea
			\alpha &\equiv& \left( \frac{\Lambda}{8\pi c_r} \right)^{\frac{1}{4}} a \;,  \label{alpha-def} \\
			\chi &\equiv& \gamma^{-1} \eta \;,  \label{chi-def} \\
			\kappa &\equiv& \gamma k \;, \label{kappa-def} \\
				\gamma^2 &\equiv& \frac{3}{\sqrt{8\pi c_r \Lambda}} \;. \label{eq:Gamma}
		\eea
    \label{scaled}
	\ees
Equation~\eqref{eq:FriedmannOriginal} can then be written
	\be\label{eq:FriedmannAlpha}	\frac{d\alpha}{d\chi}= \sqrt{1 + \alpha^4} \;. \ee
Integrating~\eqref{eq:FriedmannAlpha}  and choosing the constant of integration such that $\alpha|_{\chi=0}=1$ gives
	\be
		\alpha= \frac{ \Jacobi{cn}{2\chi\big|\frac{1}{2}} }{ \sqrt{1-\sqrt{2}\,\Jacobi{sn}{2\chi\big|\frac{1}{2}}\Jacobi{dn}
			{2\chi\big|\frac{1}{2}}} } \;,
	\ee
where $\text{sn}$, $\text{cn}$, and $\text{dn}$ are the Jacobi elliptic functions in the notation of~\cite{abramowitz_stegun_handbook_1964}.
	The limits of $\chi$ defined by $\alpha|_{\chi_0}=0$ and $\alpha|_{\chi_\infty}=\infty$ are given by
	\be	- \chi_0 = \chi_\infty =\frac{1}{2}K\!\left(\frac{1}{2}\right) =  0.927037\dots \;, \ee
where $K$ is the complete elliptic integral of the first kind.
A plot of $\alpha$ during the preinflationary era is shown in
Fig.~\ref{fig:AlphaPlot}.

The mode equation~\eqref{ModeEquation}written in terms of $\chi$ is
	\begin{equation}\label{eq:ModeEquationKappaM}
		\frac{d^2\psi_\kappa}{d\chi^2} + \left( \kappa^2 + \gamma^2 M_a^2 \right) \psi_\kappa = 0 \;.
	\end{equation}
As discussed in the introduction, in our model we compute the power spectrum using the modes of a massless minimally coupled scalar field, $m = \xi = 0$.  For this field substituting \eqref{MaDef} into \eqref{eq:ModeEquationKappaM} and using~\eqref{R-Lambda} and~\eqref{scaled} gives
	\be\label{eq:ScaledModeEquation}
		\frac{d^2\psi_\kappa }{d\chi^2}+ \left( \kappa^2 - 2\alpha^2 \right) \psi_\kappa = 0 \;.
	\ee
	The Wronskian condition in terms of scaled conformal time is
	\begin{equation}\label{eq:ScaledWronskian}
		\psi_\kappa \frac{d\psi^*_\kappa }{d\chi}- \psi^*_\kappa  \frac{d\psi_\kappa}{d\chi}= i\gamma \;.
	\end{equation}
	
	At late times $\chi\rightarrow\chi_\infty$, the spacetime is asymptotically de Sitter, and thus the mode equation~\eqref{eq:ScaledModeEquation} approaches the mode equation for pure de Sitter space.  In pure de Sitter space the solution corresponding to the Bunch-Davies state, which we will call $v_\kappa$, is
	\be\label{eq:BunchDavies}
		v_\kappa = \sqrt{\frac{\gamma}{2\kappa}} e^{-i\kappa(\chi-\chi_\infty)} \left[ 1 - \frac{i}{\kappa(\chi-\chi_\infty)}
			\right] \;.
	\ee
It satisfies the Wronskian condition~\eqref{eq:ScaledWronskian}.  It plus its complex conjugate form a set of linearly independent solutions to~\eqref{eq:ScaledModeEquation} in the limit $\chi\rightarrow\chi_\infty$.  For all times it is possible to change variables so that the general solution to the exact mode equation~\eqref{eq:ScaledModeEquation} is in the form
	\bes\label{eq:GeneralSolution}\bea
		\psi_\kappa &=& c_1(\kappa;\chi) v_\kappa + c_2 (\kappa;\chi) v_{\kappa}^{*} \;,	\label{eq:GeneralSolutionNormal}
			\\
		\frac{d\psi_\kappa}{d\chi} &=& c_1(\kappa;\chi) \frac{dv_\kappa}{d\chi} + c_2(\kappa;\chi) \frac{d v_\kappa^*}
			{d\chi} \;.\label{eq:GeneralSolutionPrime}
	\eea \ees
The coefficient functions $c_1(\chi)$ and $c_2(\chi)$ are defined by demanding that \eqref{eq:GeneralSolution} hold for all $\chi$, and are given by
	\bes\label{eq:c1c2}\bea
		c_1(\kappa;\chi) &=& -\frac{i}{\gamma} \left( \psi_\kappa \frac{d v_\kappa^*}{d\chi}- \frac{d\psi_\kappa}{d\chi}
			v_\kappa^* \right) \;,\label{eq:c1} \\
		c_2(\kappa;\chi) &=&\frac{i}{\gamma} \left( \psi_\kappa \frac{d v_\kappa}{d\chi} - \frac{d \psi_\kappa}{d \chi}
			v_\kappa \right) \;.\label{eq:c2}
	\eea\ees
As de Sitter space is approached in the limit $\chi\rightarrow\chi_\infty$, $c_1$ and $c_2$ approach constant values.
First-order differential equations for $c_1$ and $c_2$ can be obtained by using \eqref{eq:GeneralSolution} in \eqref{eq:ScaledModeEquation}:
	\bes\bea	
		\frac{dc_1}{d\chi} &=& - \frac{2i}{\gamma}\left[\frac{1}{(\chi_\infty - \chi)^2}-\alpha^2(\chi)\right]
			\left(c_1 |v_\kappa|^2 + c_2 v_\kappa^{*2}\right) \; , \label{eq:ODEc1} \\
		\frac{dc_2}{d\chi} &=& \frac{2i}{\gamma}\left[\frac{1}{(\chi_\infty - \chi)^2}-\alpha^2(\chi)\right]
			\left(c_1 v_\kappa^2 + c_2 |v_\kappa|^2\right) \; .\label{eq:ODEc2}
	\eea  \label{eq:ODEc1c2}\ees
The explicit form for the Bunch-Davies state \eqref{eq:BunchDavies} can then be substituted to get expressions where $\gamma$ cancels out.  The state for the massless minimally coupled scalar field can be specified by choosing values of $c_1$ and $c_2$ at some specific time $\chi_m$ for all values of $\kappa$.

\section{States for the massless minimally coupled scalar field}
\label{sec:States}

\subsection{Sudden approximation}
\label{sec:sudden}

The model described above is designed to give a smooth transition from an initially radiation-dominated universe to de Sitter space, such as would be expected to occur in any realistic model of inflation with an approximately homogeneous and isotropic preinflationary phase.
An even simpler approximation is to suddenly switch from a pure radiation-dominated universe to a pure de Sitter universe at some particular time.
This is called the \textit{sudden approximation.}
It has the advantage that the initial vacuum state for the massless minimally coupled scalar field is just the conformal vacuum,
\be \psi_k = \frac{1}{\sqrt{2k}}e^{-i k \eta}  \;, \label{psi-conformal} \ee
because in a pure radiation-dominated universe the scalar curvature is zero.

The matching can be done by choosing $\eta$, $a$, and $a'$ to be continuous across the matching surface.
A complete set of solutions to the mode equation~\eqref{ModeEquation} for a given value of $k$ in pure de Sitter space consists of the mode function for the Bunch-Davies state and its complex conjugate.
Multiplying each of these times a constant and matching the mode functions and their first derivatives at the sudden transition time $\eta_s$ fixes the values of the two matching coefficients.

Our model is not strictly speaking compatible with a pure radiation-dominated universe because we take $R$ to be a nonzero constant.
Similarly, because our model contains radiation, it is not strictly speaking compatible with pure de Sitter space. Nevertheless it is possible to use our model to obtain the same results for the power spectrum as one gets from the sudden approximation.  This is useful so that we can directly compare the resulting power spectrum with those of previous calculations as well as with our results for adiabatic vacuum states, which are described below.

Because the matching in the real spacetime is just done at a particular value of the time $\eta=\eta_s$, there is nothing to prevent one from using \eqref{chi-def} and~\eqref{kappa-def} for some fixed value of $c_r \,\Lambda$ to change from $\eta$ to $\chi$ and from $k$ to $\kappa$.
After doing so, the matching equations are equivalent to \eqref{eq:c1c2} evaluated at the matching time $\chi_s$.
However, for  $\chi > \chi_s$ the spacetime is pure de Sitter space, so instead of being starting values for a numerical integration, in the sudden approximation
$c_1$ and $c_2$ are fixed constants.

In terms of the scaled coordinates, \eqref{psi-conformal} becomes
\begin{equation}
	\psi_\kappa = \sqrt{\frac{\gamma}{2 \kappa}} e^{-i \kappa \chi} \;. \label{psi-conformal-kappa}
\end{equation}
The Bunch-Davies state in these coordinates in pure de Sitter space is given in \eqref{eq:BunchDavies}.
Substituting into \eqref{eq:c1c2} at $\chi=\chi_s$ gives
	\bes\bea
			c_{1s} &=&\left[1+\frac{i}{\kappa(\chi_s-\chi_\infty)}
				-\frac{1}{2\kappa^2(\chi_s-\chi_\infty)^2}\right] e^{-i\kappa\chi_\infty} \; ,\\
			c_{2s} &=& -\frac{1}{2\kappa^2(\chi_s-\chi_\infty)^2}e^{i\kappa(\chi_\infty - 2\chi_s)} \;.
	\eea\label{c1-c2-sudden}\ees

The sudden approximation is an extreme limit and often results in a state that is not physically acceptable.
As shown below, that is the case here.
It is for this reason that it is useful to consider a model in which the Universe evolves continuously from a radiation-dominated era to the inflationary era such as the one described in Sec.~\ref{sec:OurModel}.
As shown in Sec.~\ref{sec:natvacstate}, we have not found a physically acceptable natural initial vacuum state for the massless minimally coupled scalar field in this model.
In lieu of a natural initial vacuum state, it is often useful to consider various adiabatic vacuum states, and these are discussed next.

\subsection{Adiabatic vacuum states}\label{sec:Adiabatic}

A choice of vacuum state for our model can be made by specifying the values of $c_1$ and $c_2$ in \eqref{eq:c1c2} at some time $\chi_m$ for each value of $\kappa$.
The solutions to the mode equations can then be obtained at any other time by numerically integrating \eqref{eq:ODEc1c2} forward (or backward) in time.
In this section we discuss adiabatic vacuum states.
These are exact states for the quantum field that are specified by using a WKB approximation to provide starting values for the modes and their first time derivatives at some particular time that we call the matching time~\cite{parker-thesis,pf,fp,fph,bunch}.

To understand how the WKB expansion works for the scaled coordinates,
it is useful to begin with the original coordinates and the original form of the mode equation~\eqref{ModeEquation}.
For the massless minimally coupled scalar field, the mode equation can be written in the form
\begin{equation}
	\psi^{''}_k + \left(k^2 - \frac16 a^2 R  \right)\psi_k = 0 \;.  \label{psi-mmc0}
\end{equation}
Note that for the model we are using, $R = 4 \Lambda$.
Then one makes the change of variable
\begin{equation}
	\psi_k = \frac{1}{\sqrt{2W_k}} \exp\left[-i\int_{\eta_1}^\eta W_k(\eta') d\eta' \right] \; , \label{wkb-orig}
\end{equation}
where $\eta_1$ is an arbitrary constant.  This automatically ensures the Wronskian condition \eqref{Wronskian}, with the result that
	\be W^2 = k^2 - \frac{a''}{a} - \left( \frac{W''}{2W}  - \frac{3W^{\prime 2}}{4W^2}\right)  \; .\ee
One starts with zeroth order in terms of derivatives of the metric and then iterates.
At each iteration the new terms contain two more derivatives of the scale factor than the previous ones.
Thus
	\bes\bea
		W^{(0)} &=& k \; , \\
		W^{(2)} &=&  k - \frac{a''}{2 k a} \;, \\
		W^{(4)} &=& k - \frac{a''}{2 k a} + \frac{2a^{\prime 2} a'' - 2 a a' a''' -2 a a^{\prime\prime 2}
			+ a^2 a^{\prime\prime\prime\prime}}{8 k^3 a^3} \; .
	\eea\ees
	
An adiabatic state is an exact state for the quantum field that is obtained by using the WKB approximation to some order to fix the starting values for the modes at some particular matching time $\eta_m$.
One does this by substituting the expression for $W$ at some order into \eqref{wkb-orig} and equating with the exact mode function.
One does the same for the first time derivative of \eqref{wkb-orig}.
For a given order there are many possible adiabatic states, in part because one obtains different states for different matching times.

For the specific model we are considering and the scaled variables that we are using, one can think of $\alpha^2$ as being of second adiabatic order and each derivative of $\alpha$ then giving an extra adiabatic order.
The reason is that, as mentioned above, the scalar curvature $R$ is a constant for our model and as is seen from Eq.~\eqref{psi-mmc0} is multiplied by a factor of $a^2$ in the mode equation.
Then the WKB approximation in terms of scaled variables is
\begin{equation}\label{eq:AdiabaticStates}
	\psi_\kappa = \sqrt{\frac{\gamma}{2W}} \exp \left[-i\int_{\chi_1}^\chi W(\chi') d\chi' \right]
\end{equation}
where
	\be
		W^2 = \kappa^2-2\alpha^2 - \left[ \frac{1}{2W}\frac{d^2W}{d\chi^2} - \frac{3}{4W^2}\left(\frac{dW}{d\chi}
			\right)^2 \right] \;.
	\ee
One easily finds that
	\bes	\bea
		W^{(0)} &=& \kappa \;, \label{W-ad-0} \\
		 W^{(2)} &=& \kappa - \frac{\alpha^2}{\kappa} \;,  \label{W-ad-2} \\
		 W^{(4)} &=& \kappa - \frac{\alpha^2}{\kappa} +\frac{2\alpha^4+1}{2\kappa^3}  \;,   \label{W-ad-4}
	\eea \label{W-ad} \ees
where we used \eqref{eq:FriedmannAlpha} to eliminate all the derivatives of $\alpha$.

To do the adiabatic matching at a time $\chi_m$, it is easiest to choose the lower limit of the integration variable to be $\chi_1 = \chi_m$.
Then
	\bea
		\psi^W_\kappa(\chi_m) &=& \sqrt{\frac{\gamma}{2 W}} \; , \nonumber \\
       	\left(\frac{d}{d \chi}{\psi^W_\kappa}\right)_{\chi_m} &=& -i \sqrt{\frac{\gamma W}{2}}
       		- \frac{\sqrt\gamma}{(2 W)^{3/2}} \frac{d W}{d \chi} \;. \label{psi-wkb}
	\eea
One can compute these to a particular adiabatic order by substituting for $W$.
Strictly speaking, all that is necessary for the derivative of $W$ is to use the previous adiabatic order, although it is permissible to use the same adiabatic order.
The adiabatic state that is generated will be different in the two cases.
The result is then substituted into Eq.~\eqref{eq:c1c2} to obtain values for $c_1$ and $c_2$ at the time $\chi_m$.
This fixes the solutions to those equations.

It is important to note that the WKB approximation breaks down in our model for $\kappa \le \sqrt{2}\, \alpha$.
 Thus for any given matching time $\chi_m$ there will be values of $\kappa$ that cannot reasonably be fixed using adiabatic matching.
If one wishes to compute the stress-energy tensor for the quantum field, then it will be important to find acceptable starting values for such modes.
However, for the purposes of the power spectrum, it is in most cases sufficient to restrict attention to $\kappa > \sqrt{2} \,\alpha(\chi_m)$  for matching times $\chi_m$ that occur near the time when the semiclassical approximation becomes valid in the radiation-dominated preinflationary phase. Exceptions can occur if the
horizon size at the time of the onset of inflation, when scaled to the current time, is  significantly smaller than the horizon size today.

\section{Power Spectrum}\label{sec:PowerSpectrum}
The standard power spectrum for the field $\phi$ given in terms of wave number $k$ and conformal time $\eta$ is~\cite{anderson_short_2005}
\begin{equation}
	P_\phi(k;\eta) = \frac{k^3}{2\pi^2} \left| \phi_k(\eta) \right|^2 \;.
\end{equation}
Using~\eqref{PsiDef} and the scaled variables~\eqref{eq:ScaledVariables} gives
\begin{equation}
	P_\phi(\kappa;\chi)
	= \frac{\kappa^3 H_\Lambda^2}{2\pi^2 \gamma \alpha^2} \left| \psi_\kappa(\chi) \right|^2 \;, \label{power-spectrum-scaled}
\end{equation}
where $H_\Lambda^2 \equiv \frac{1}{3}\Lambda$.
Evaluating \eqref{power-spectrum-scaled} in the limit $\chi\rightarrow\chi_\infty$ using \eqref{eq:BunchDavies} and \eqref{eq:GeneralSolutionNormal}, one finds
\begin{equation}\label{eq:PowerSpectrumFinal}
		P_\phi(\kappa) = \frac{H_\Lambda^2}{4\pi^2} \left| c_1(\kappa;\chi_\infty) - c_2(\kappa;\chi_\infty) \right|^2 \;.
\end{equation}
The problem of calculating the late time power spectrum therefore reduces to finding $c_1(\kappa;\chi_\infty)$ and $c_2(\kappa;\chi_\infty)$.
For the model we are considering this can be accomplished by solving \eqref{eq:ODEc1c2}.

Models of the early Universe are heavily constrained by observations of the cosmic microwave background (CMB) as well as measurements of large-scale structure.
Variations in the CMB are described in terms of the parameters $C_\ell$, which are related to the power spectrum by~\cite{Riotto2002_Inflation}
	\be\label{eq:AngularPowerSpectrumGeneral}
		C_\ell = \frac{4\pi}{9}\int_0^\infty \frac{dk}{k} P_\phi( k ) j_\ell^2( k \eta_\text{eff} )
		= \frac{4\pi}{9}\int_0^\infty \frac{d\kappa}{\kappa} P_\phi( \kappa ) j_\ell^2( s^{-1}\kappa) \;.
	\ee
Here $j_\ell$ is a spherical Bessel function, $\eta_\text{eff} = \frac{r_\text{eff}}{a_0}$, where $r_\text{eff}$ and $a_0$ are the physical size of the effective horizon and scale factor today, and $s = \frac{\gamma a_0}{r_{\text{eff}}}$.
We define $a_i$ as the scale factor at the start of inflation, when the radiation and cosmological constant contributions to \eqref{eq:FriedmannOriginal} are equal.
Then using Eq.~\eqref{eq:Gamma} and $H_\Lambda^2=\frac{\Lambda}{3}$, we find $a_i = H_\Lambda^{-1} \gamma^{-1}$, and therefore
\begin{equation}\label{eq:SDefinition}
	s = \left( \frac{a_0}{a_i} \right) \left( \frac{H^{-1}_\Lambda}{r_\text{eff}} \right) \;.
\end{equation}
This means that $s$ corresponds approximately to the ratio of the size of the horizon at the start of inflation, scaled to the current time, to the effective horizon today.

\section{Results}
\label{sec:Results}

\begin{figure}[h]
	\includegraphics[width=\linewidth]{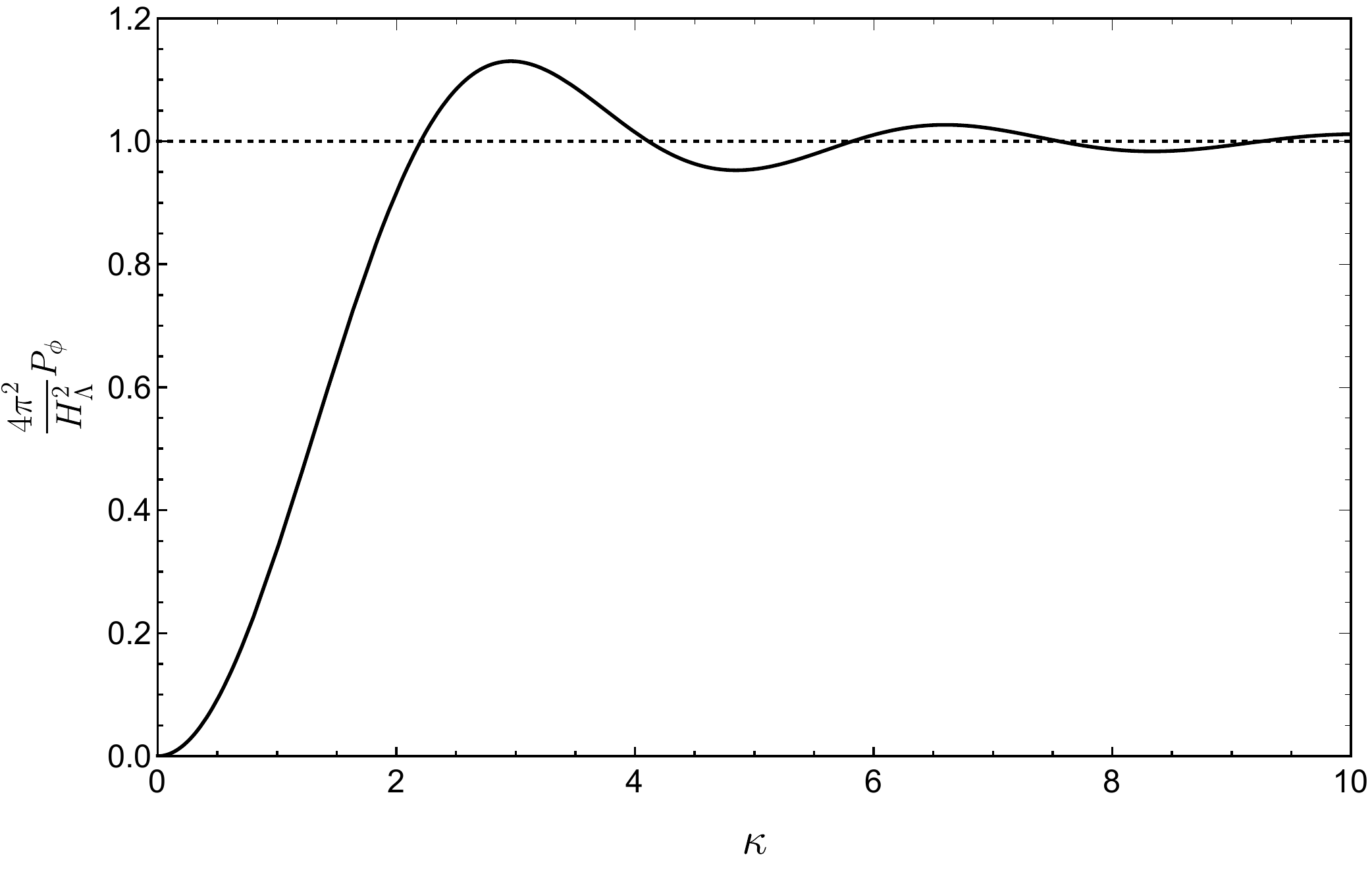}
	\caption{Power spectrum from a sudden approximation with the transition occurring at $\chi_s = 0$. Enhancements in the power spectrum are seen compared to the Bunch-Davies value, represented by the dotted line. \label{fig:Sudden}}
\end{figure}

Although it is not always obvious exactly how a state for the massless minimally coupled scalar field was chosen, it seems likely that many previous calculations of the power spectrum~\cite{hirai_initial_2003,cline_does_2003,hirai_initial_2005,hirai_effect_2006,powell_pre-inflationary-2007,hirai_effect_2007,marozzi_on_2011,das_revisiting_2015} for a radiation-dominated preinflationary phase made use of either the sudden approximation or something very similar to it.
Therefore it is useful to begin with the power spectrum we obtain for the sudden approximation.
Substituting~\eqref{c1-c2-sudden} into~\eqref{eq:PowerSpectrumFinal} one finds
\begin{multline}
	P_\phi = \frac{H_\Lambda^2}{4\pi^2} + \frac{H_\Lambda^2}{8\pi^2 \kappa^4 (\chi_\infty - \chi_s)^4}
	\bigl\{1 + \left[2 \kappa^2(\chi_\infty - \chi_s)^2 - 1 \right] \cos \left[ 2\kappa (\chi_\infty - \chi_s)\right]
	\\
	{}- 2 \kappa (\chi_\infty - \chi_s) \sin \left[ 2 \kappa (\chi_\infty - \chi_s) \right] \bigr\}\; .
\end{multline}
This spectrum oscillates and has a peak value about a factor of $1.13$ times the Bunch-Davies constant value, independent of the time of the sudden transition $\chi_s$.
The resulting spectrum is shown in Fig.~\ref{fig:Sudden} for the choice $\chi_s=0$.
The large oscillations and enhanced values compared with the power spectrum for the Bunch-Davies state are qualitatively identical with most previous results in the literature.

\begin{figure}[h]
	\includegraphics[width=\linewidth]{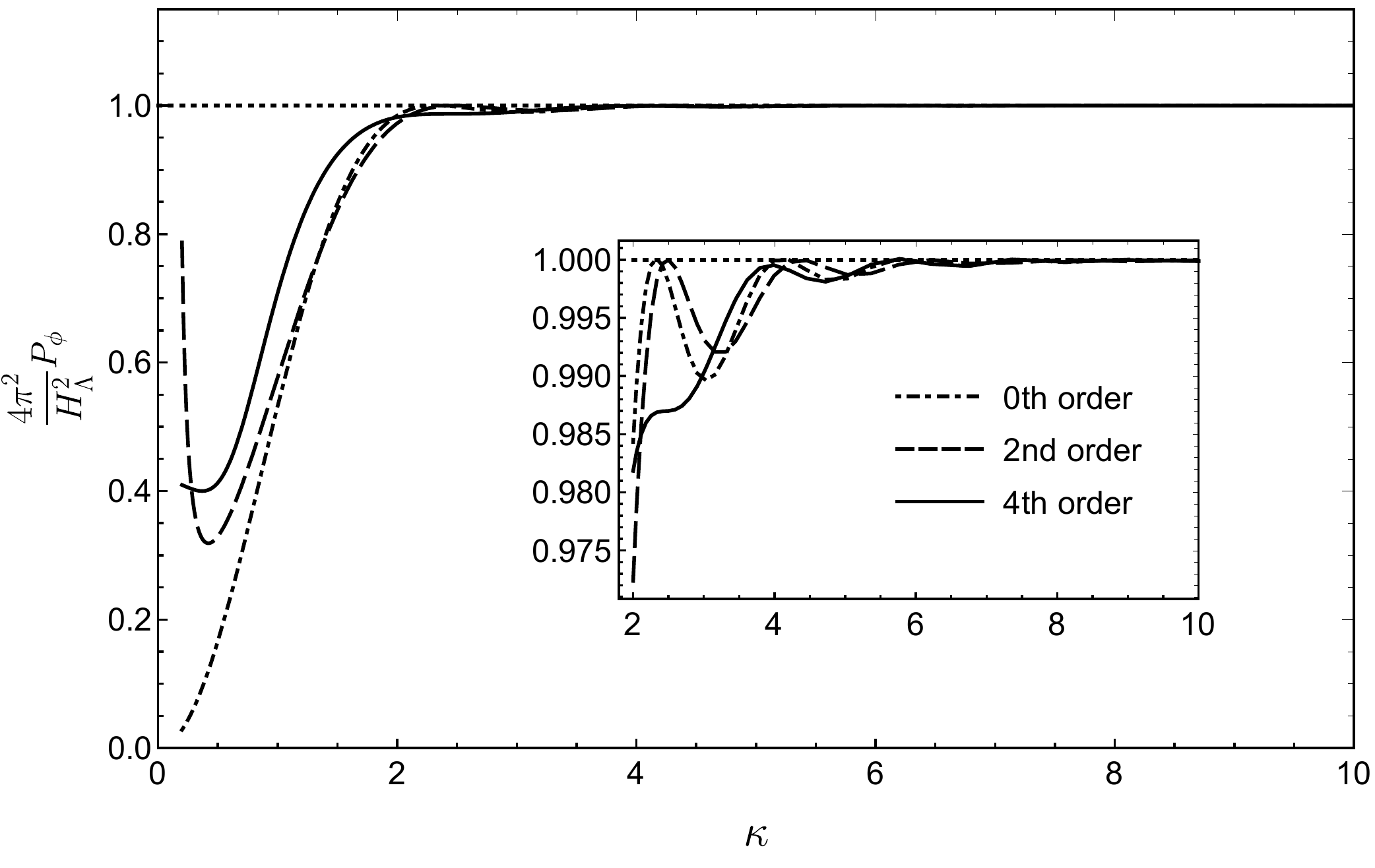}
	\caption{Comparison of the power spectra for adiabatic vacuum states of order zero, two, and four, when the matching is done at $\alpha_m = 0.1$ ($\chi_m=-0.827$).  Note that all orders show oscillatory
		behavior, but	this behavior is much smaller at higher orders.  Note also that the power spectrum never noticeably exceeds 1, the
		Bunch-Davies value.\label{fig:AdiabaticPowerSpectrum}}
\end{figure}
\begin{figure}[h]
	\includegraphics[width=\linewidth]{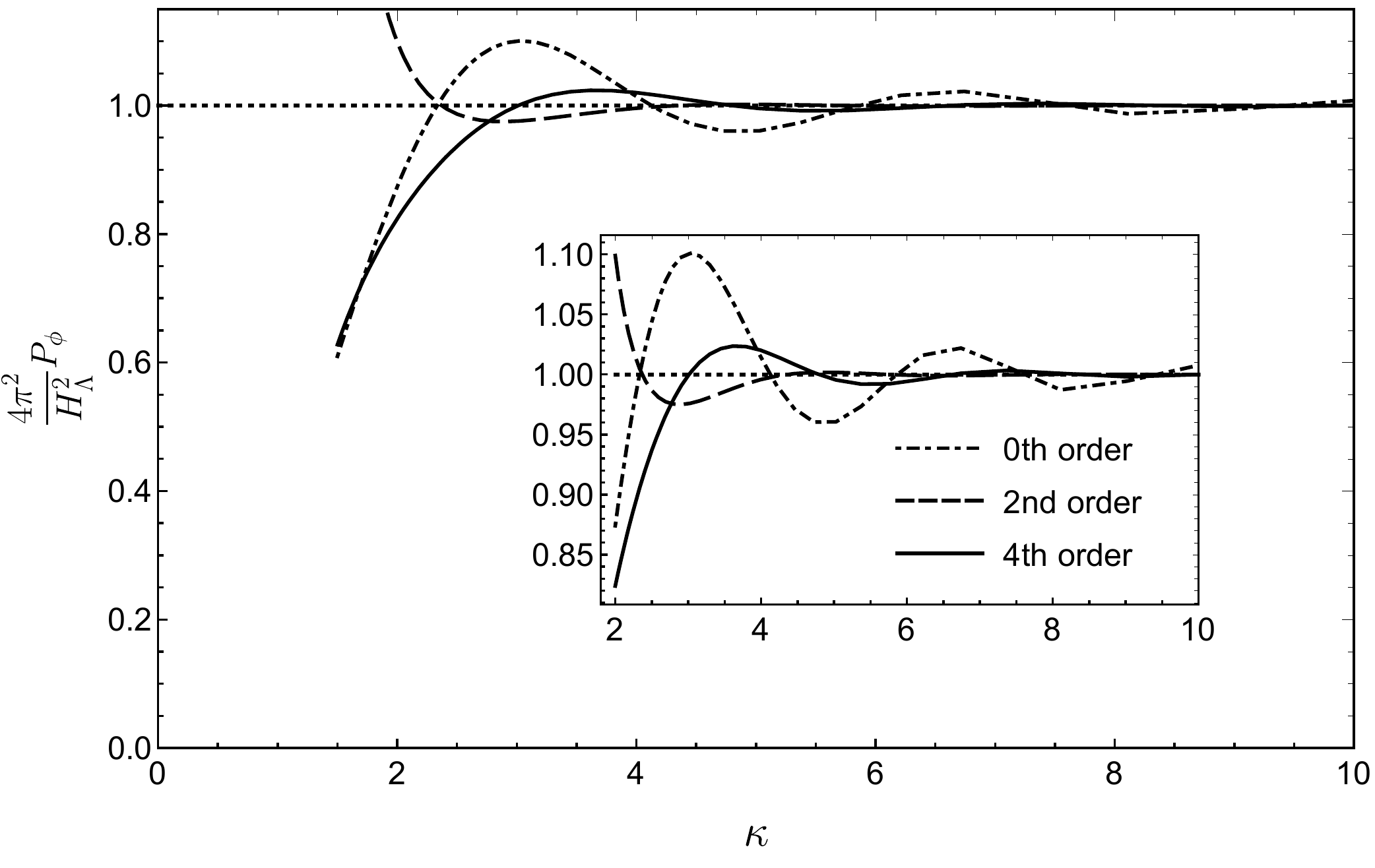}
	\caption{Comparison of the power spectra for adiabatic vacuum states of order zero, two, and four, when the matching is done at $\alpha_m=1$ ($\chi_m=0$).\label{fig:AdiabaticPowerSpectrum2}}
\end{figure}

For the adiabatic vacuum states we considered, starting values for $c_1$ and $c_2$ were calculated at the matching time $\chi_m$ as discussed above, and then Eqs.~\eqref{eq:ODEc1c2} were solved numerically to find their asymptotic values.
These asymptotic values were then substituted into \eqref{eq:PowerSpectrumFinal} to obtain the power spectrum.
Our results for the matching time when $\alpha = 0.1$ are shown in Fig.~\ref{fig:AdiabaticPowerSpectrum}.
Note that if inflation occurs at the GUT scale, then since $\alpha=1$ is the onset of inflation in our model, the energy at the matching time is 10 times larger than the energy at the onset of inflation.
However, the energy at the matching time is still 100 times smaller than the Planck scale, so this is a conservative estimate of when the semiclassical approximation first became valid.
For comparison purposes, Fig.~\ref{fig:AdiabaticPowerSpectrum2} shows our results for a matching time corresponding to the onset of inflation when $\alpha = 1$.
This is too late to be a natural time for the matching but provides an important illustration of how much the adiabatic vacuum states depend on the matching time.

As can be seen from the inset in Fig.~\ref{fig:AdiabaticPowerSpectrum}, where the matching is done at $\alpha = 0.1$, all three adiabatic orders shown (zeroth, second, and fourth) have a small amount of oscillatory behavior, with the amplitudes of the oscillations decreasing as the adiabatic order increases.
Note that in no case is there a noticeable enhancement of the power spectrum above the standard Bunch-Davies state.
In contrast, using a matching time when $\alpha = 1$, Fig.~\ref{fig:AdiabaticPowerSpectrum2} shows that the oscillations in the power spectra are significantly enhanced in comparison with the earlier matching time for a given adiabatic order.

The one case we are aware of where the power spectrum was computed using adiabatic states for the specific cosmological model we used was in Ref.~\cite{wang_effects_2008}.
There, the power spectrum was computed for zeroth-order adiabatic states with various matching times.
For relatively late matching times, our computations of the power spectrum for zeroth-order adiabatic vacuum states agree qualitatively with theirs.
However, we do find numerical differences when the matching is done at late times that we cannot explain.
We also find a qualitative difference when the matching is done at early times in that we always see oscillations in the power spectrum whereas their results show monotonic behavior.

In~\cite{Nicholson2008_The}, the power spectrum was computed numerically for a radiation-dominated preinflationary phase in the context of slow-roll inflation.
The authors assumed a zeroth-order adiabatic state.
It is unclear what matching time they used, but if it was near the onset of inflation, then our results agree qualitatively with theirs.

\begin{figure}[h]
	\includegraphics[width=\linewidth]{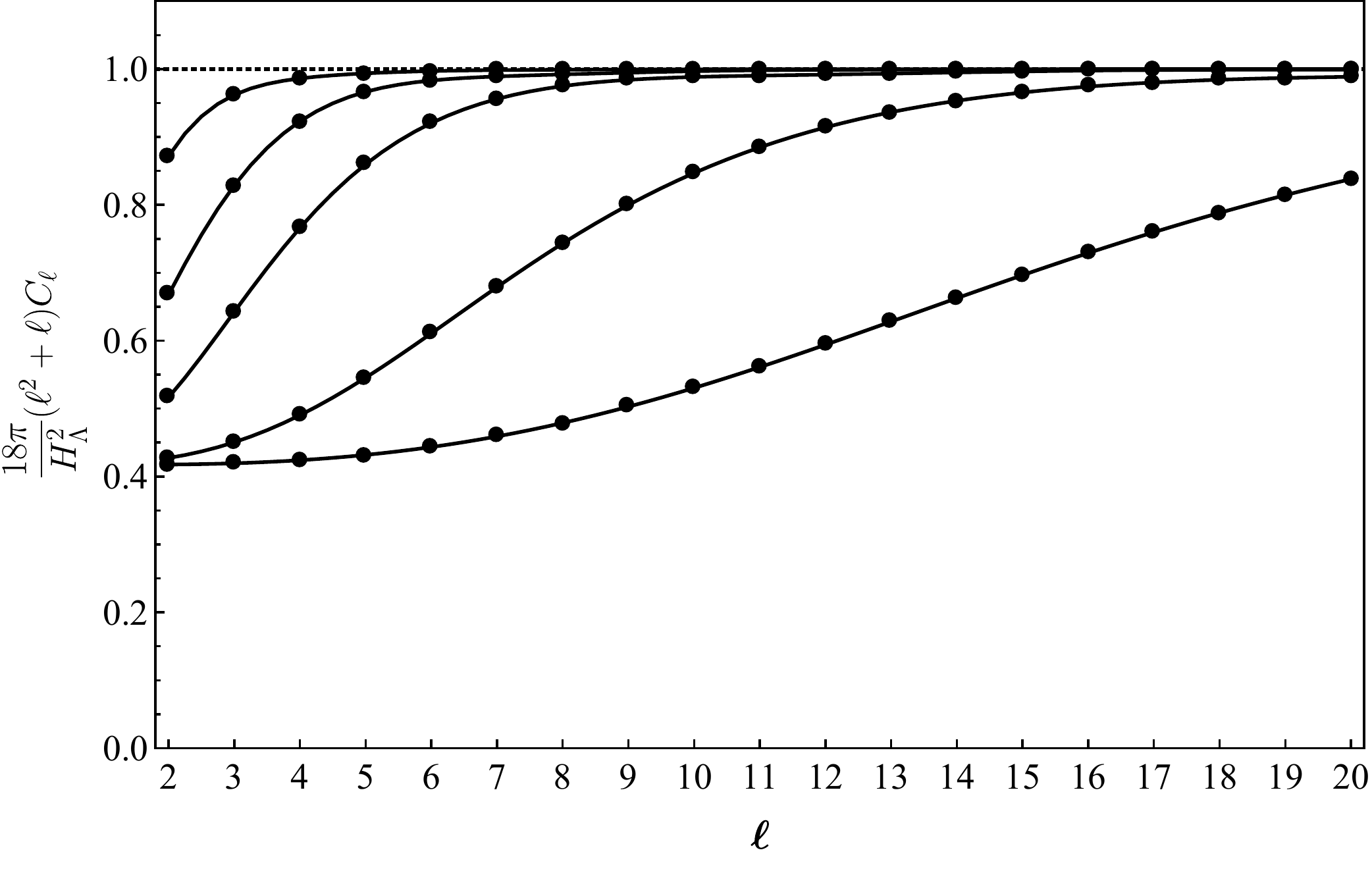}
	\caption{Contributions to the angular power spectrum from modes with $\kappa \ge \sqrt{2} \alpha_m$ for a fourth-order adiabatic vacuum state with the adiabatic
matching done at $\alpha_m=0.1$.    From top to bottom, the curves are for $s=0.50$, 0.30, 0.20, 0.10, and 0.05.
 \label{fig:AngularPower4th}}
\end{figure}

\begin{figure}[h]
	\includegraphics[width=\linewidth]{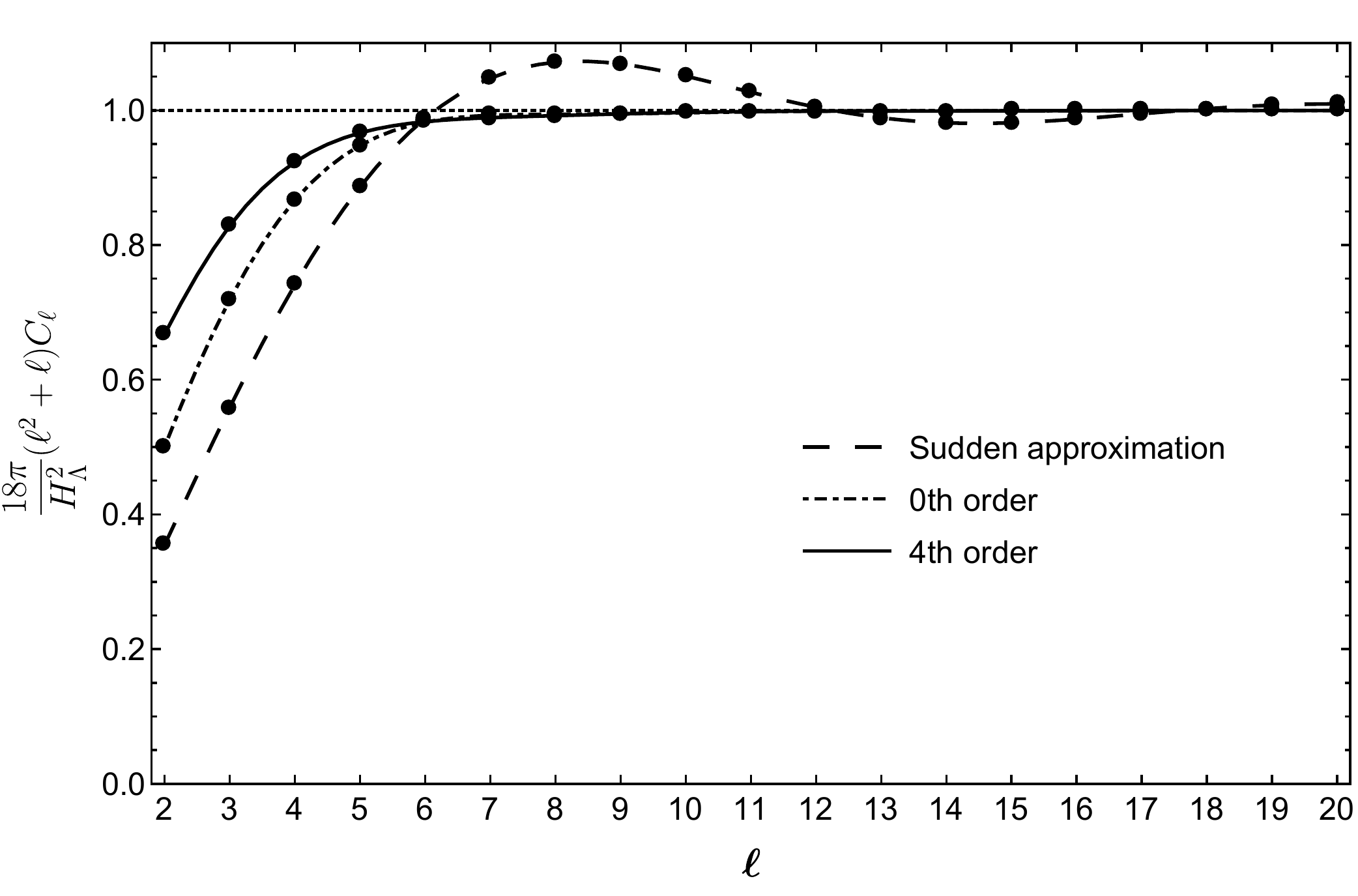}
	\caption{Angular power spectrum for the sudden approximation
and for the contributions from modes with $\kappa \ge \sqrt{2} \alpha_m$ to the power spectrum for zeroth- and fourth-order adiabatic vacuum states.   For the sudden approximation the matching is done at $\alpha_s = 1$, while for the adiabatic vacuum states the
matching was done at $\alpha_m=0.1$. In each case, $s=0.3$.  \label{fig:AngularPowerSpectrum30}}
\end{figure}

Once the power spectrum has been computed, the angular power spectrum can be calculated using \eqref{eq:AngularPowerSpectrumGeneral}.
Figure~\ref{fig:AngularPower4th} shows the resulting spectrum for a fourth-order adiabatic state and a matching time of $\alpha = 0.1$.
It can be shown that in the limit $s \to \infty$ the resulting angular power spectrum is flat and, thus, independent of $\ell$, but if $s$ is not too large, there is a suppression of the angular power spectrum for small $\ell$.
Note that for matching at an early time such as $\alpha = 0.1$, any suppression of the $\ell=2$ component is accompanied by a comparable but smaller suppression of $\ell=3$ and other small $\ell$ values.
Figure~\ref{fig:AngularPowerSpectrum30} compares the results for the zeroth- and fourth-order adiabatic states with matching at $\alpha = 0.1$ with the sudden approximation for the case $s = 0.3$.

\subsection{Adiabatic states in de Sitter space}
\label{sec:WKB-dS}

To understand why the oscillations for a given matching time are smaller for larger adiabatic orders, it is useful to switch to pure de Sitter space where analytic solutions to the mode equation are known and the power spectrum for any state can be computed analytically.
For simplicity, we will also revert to unscaled variables, so that
\begin{subequations}
	\begin{align}
		\begin{split}
			a_\text{dS} &= -\frac{1}{H\eta} \;,
		\end{split}\\
		\begin{split}
			v_k &= \frac{e^{-ik\eta}}{\sqrt{2k}} \left(1-\frac{i}{k\eta}\right) \; ,
		\end{split}\\
		\begin{split}
			\psi_k &= \frac{1}{\sqrt{2W}} e^{-i\int W d\eta} \; ,
		\end{split}\\
		\begin{split}
			W^2 &= k^2-\frac{a_{\text{dS}}''}{a_{\text{dS}}} - \left(\frac{W''}{2W} - \frac{3W^{\prime 2}}{4W^2}\right) \;
		\end{split}
	\end{align}
\end{subequations}
We can find $c_1$ and $c_2$ directly from Eq.~\eqref{eq:c1c2}, which become
\begin{subequations}
	\begin{align}
		\begin{split}
			c_1 &= -i\left(\psi_k v_k^{*\prime} - \psi_k' v_k^* \right) \; , \label{c1Init}
		\end{split}\\
		\begin{split}
			c_2 &= i\left(\psi_k v_k^{\prime} - \psi_k' v_k \right) \;. \label{c2Init}
		\end{split}
	\end{align}
\end{subequations}
Recall that primes denote derivatives with respect to $\eta$.
The power spectrum is given by Eq.~\eqref{eq:PowerSpectrumFinal}, but $c_1$ and $c_2$ are now independent of time
\begin{equation}
	P_\phi = \frac{H^2}{4\pi^2} \left| c_1(k) - c_2(k) \right|^2 \;.
\end{equation}
The power spectra for an adiabatic matching time $\eta_m$ for zeroth-, second-, and fourth-order adiabatic states are
\begin{subequations}\label{eq:deSitterPowerSpectra}
	\begin{align}
	\begin{split}
		P_\phi^{(0)} =& \frac{H^2}{4\pi^2} + \frac{H^2}{8\pi^2 k^4 \eta_m^4} \left[ 1 + \left( 2k^2\eta_m^2-1\right)\cos(2 k \eta_m) - 2 k \eta \sin(2 k \eta_m) \right] \; ,
	\end{split}\\
	\begin{split}
		P_\phi^{(2)} =& \frac{H^2}{4\pi^2} + \frac{H^2}{8 \pi^2 k^4 \eta_m^4 (k^2\eta_m^2-1)^3}
			\big[ k^2 \eta_m^2+ 1
	\end{split}\nonumber\\
	\begin{split}
		&- (2 k^6 \eta_m^6- 2 k^4 \eta_m^4 -k^2\eta_m^2 +1 )\cos(2 k \eta_m) - 2 k \eta_m \sin(2 k \eta_m) \big] \; ,
	\end{split}\\
	\begin{split}
		P_\phi^{(4)} =& \frac{H^2}{4\pi^2} + \frac{H^2}{8\pi^2k^6\eta_m^6(k^4\eta_m^4-k^2\eta_m^2+1)^3} \bigl[ k^6 \eta_m^6 + 6k^4\eta_m^4-3k^2\eta_m^2 +1
	\end{split}\nonumber\\
	\begin{split}
		&+(2k^{12}\eta_m^{12}+2k^{10}\eta_m^{10}-8k^8\eta_m^8+17k^6\eta_m^6-14k^4\eta_m^4+5k^2\eta_m^2-1)
			\cos(2 k \eta_m)
	\end{split}\nonumber\\
	\begin{split}
		&+(2k^{11}\eta_m^{11}-8k^9\eta_m^9+14k^7\eta_m^7-20k^5\eta_m^5+ 8k^3\eta_m^3-2k\eta_m) \sin(2 k \eta_m) \bigr] \; .
	\end{split}
	\end{align}
\end{subequations}
For a universe with a preinflationary radiation-dominated era, de Sitter space is approached in the late time limit.
Therefore, the terms in Eq.~\eqref{eq:deSitterPowerSpectra} that provide the leading-order behavior in our model are those with the largest power of $k\eta_m$.
We find that the oscillatory terms always dominate the nonoscillatory terms.
Furthermore, the leading-order oscillatory terms have smaller and smaller contributions at higher order, with zeroth order being $(k\eta_m)^{-2}$, second order $(k\eta_m)^{-4}$, fourth order $(k\eta_m)^{-6}$, and so on.
That is, oscillatory terms will contribute the most at zeroth order and then contribute less and less as the order is increased.

\subsection{Discussion}\label{sec:Discussion}

As can be seen in Figs.~\ref{fig:AdiabaticPowerSpectrum} and~\ref{fig:AngularPowerSpectrum30}, for adiabatic matching at a relatively early time there is a suppression of the power spectrum at small wave numbers $\kappa$ compared to the spectrum of the Bunch-Davies state.
In contrast, for a sudden approximation we find an enhancement at certain values of $\kappa$, as seen in Figs.~\ref{fig:Sudden} and~\ref{fig:AngularPowerSpectrum30}.
The latter behavior agrees qualitatively with the results in~\cite{cline_does_2003,powell_pre-inflationary-2007,marozzi_on_2011,das_revisiting_2015}, where it appears that some type of sudden approximation was used.  The sudden approximation was also used and an enhancement of the power spectrum for certain values of the momentum parameter was also observed in~\cite{hirai_initial_2003,hirai_initial_2005,hirai_effect_2006,hirai_effect_2007}.  However, our results disagree in detail with theirs.
The qualitative differences between the power spectrum for the sudden approximation and for states with adiabatic matching at an early time imply that it is necessary to not use a sudden approximation but instead to work in a spacetime where there is a smooth transition to the inflationary era and compute the power spectrum using solutions to the exact mode equation in that spacetime, which in most cases will be obtained numerically.
This point was made in~\cite{wang_effects_2008}, where, as discussed above, the power spectrum was computed for zeroth-order adiabatic vacuum states at various matching times for the same model that we consider here.

For large values of $\kappa$ the power spectrum approaches the constant value it has for the Bunch-Davies state.
That in turn, gives an approximately flat spectrum at large $\ell$, as shown in Fig.~\ref{fig:AngularPower4th}.  A more realistic model would include not just a simple inflation era, but rather an evolving scalar field in an appropriate potential, which would presumably result in a tilted spectrum.
This spectrum would then have to be processed through all of the subsequent stages of cosmology to reproduce the CMB that we observe.
We think it is likely that our findings of the suppression of small $\ell$ with no enhancement at any $\ell$ would persist in a more realistic model of this type.

The suppression of the spectrum at small $\ell$ could provide one explanation of the anomalously small value of the quadrupole moment of the CMB~\cite{Bennet2003_First}.
In~\cite{wang_effects_2008}, where the same model that we are using was investigated, this suppression at small values of $\ell$ was observed.
By examining Fig.~\ref{fig:AngularPower4th}, we see that any significant suppression of the $\ell=2$ modes comes at the expense of also suppressing $\ell=3$, which is not observed.
It should also be noted that values which lead to significant suppression of $\ell=2$ have relatively small $s$ values, such as $s \approx 0.3$.
Recall that $s$ corresponds roughly to the ratio of the size of the horizon at the start of inflation, scaled to the current time, to the effective horizon today (see~\eqref{eq:SDefinition}).
This means that for $s \lesssim 1$ we cannot explain the homogeneity and flatness of the current Universe exclusively in terms of inflation; there must be some mechanism which makes the Universe uniform on scales slightly larger than the horizon size at the start of inflation.  However, because of the existence of an inflationary phase, the flatness of the Universe today does not require the ultrafine-tuning that would typically be needed without inflation.
It would only take a very modest fine-tuning at the start of inflation to result in an approximately flat Universe today.

It is worth noting that the trans-Planckian censorship conjecture (TCC)~\cite{bedroya2019transplanckian} which states that trans-Planckian modes should never cross the horizon in an expanding universe and become classical, provides a restriction on the duration of the inflationary phase~\cite{Bedroya_2020}.  In~\cite{berera2020role} it was argued that it is possible to modify the TCC in ways that allow trans-Plankian modes to cross the horizon.   However, with these modifications the TCC still provides a restriction on the duration of inflation.

\section{Summary and Conclusions}
\label{sec:Conclusions}

In Sec.~\ref{sec:Prediction} it was argued that if there was a preinflationary era in which the semiclassical approximation in gravity was valid and if the Universe, or our part of it, was approximately homogeneous and isotropic during that time, then it is very likely the Universe expanded like a radiation-dominated universe during that era.  The
argument is based on several assumptions and a proof that is given in the Appendix.  The assumptions are
that (i) the dominant quantum fields should behave like free fields during that epoch; (ii)  higher derivative terms necessary for the renormalization of the stress-energy tensor should be small since the semiclassical approximation is assumed to be valid; and (iii) massive spin-$\frac{1}{2}$ and spin-$1$ fields can be modeled using massive conformally coupled spin-$0$ fields.  The proof shows that, to leading order in the limit that the scale factor vanishes, the energy density of a massive conformally coupled scalar field is of the same form as that of classical radiation for a very large class of physically acceptable homogeneous and isotropic vacuum states provided that $(a^{2})^{\prime}$ is finite at the initial singularity and $\int_{\eta_0}^{\eta} |[a^2(x)]''| dx$ is also finite.

It was further shown that there is a state for a scalar field with arbitrary curvature coupling that goes like $\psi = (2k)^{-1/2} e^{-i k \eta}$ when the scale factor is small, so long as in the limit that the scale factor vanishes the effective mass term in the mode equation also vanishes.
This would seem to be a natural initial vacuum state because the mode equation approaches that of the conformally invariant scalar field in this limit.
It was also shown that this state is an infinite-order adiabatic vacuum state if the scale factor vanishes in the limit $\eta \to - \infty$, such as happens in pure de Sitter space, where the vacuum state is the Bunch-Davies state.
However, if the scale factor vanishes at a finite value $\eta_0$ of the conformal time and one or more of the derivatives of the effective mass term is nonzero at $\eta_0$, then this state is only a finite-order adiabatic vacuum state.  In particular, if the Universe expanded like a radiation-dominated universe near $\eta = \eta_0$, then
for an arbitrarily coupled massive scalar field the state is at most a first-order adiabatic vacuum state.
For a massless nonconformally coupled scalar field the state can be of adiabatic order two or higher in the radiation-dominated case depending on the detailed behavior of the scale factor near $\eta_0$.
For the specific model considered here it is at most a third-order adiabatic vacuum state.
For the conformally coupled massive scalar field a zeroth-order adiabatic state is enough to give a finite stress-energy tensor in a homogeneous and isotropic spacetime but only for exact homogeneity and isotropy.
For nonconformally coupled scalar fields, both massive and massless, a fourth-order adiabatic state is required for the renormalized stress-energy tensor to be finite.
It is important to point out that this does not prevent the solution from being used for small and intermediate values of the momentum parameter $k$, but it does prevent it from being used for arbitrarily large values of $k$.

In Sec.~\ref{sec:PowerSpectrum} we studied the effects of a radiation-dominated preinflationary phase on the power spectrum that is computed using the massless minimally coupled scalar field.  To do so we used a simple model in which the classical Einstein equations are solved when classical radiation and a positive cosmological constant are present.  The mode equation was solved for this model for several different states for the quantum field, and the solutions were used to compute the power spectra for these states.

We found that a sudden approximation in which the metric is exactly that of a radiation-dominated universe up to a transition time and is exactly that of de Sitter space afterward gives relatively large oscillations in the power spectrum when it is plotted as a function of the momentum parameter $k$.
For an early matching time, well before the onset of inflation, a zeroth-order adiabatic state such as the one mentioned above gives oscillations with a smaller amplitude, while higher-order adiabatic states at the same matching time give successively smaller amplitude oscillations.
Previous investigations~\cite{hirai_initial_2003,cline_does_2003,hirai_initial_2005,hirai_effect_2006,powell_pre-inflationary-2007,hirai_effect_2007,marozzi_on_2011,das_revisiting_2015} in which there is a radiation-dominated preinflationary phase have found similar results.  Those investigations appear to have made use of either a sudden approximation or zeroth-order adiabatic states.  In some cases there are disagreements with our results that are discussed in Sec.~\ref{sec:Discussion}.  All of these calculations, including ours for the better-behaved second- and fourth-order adiabatic states, predict that if inflation did not go on for too long, then there are potentially observable differences in the power spectrum from that of the Bunch-Davies state in de Sitter space.  These differences, if observed, would have the potential to give us information about the initial state of the matter fields in the Universe if there was a preinflationary radiation-dominated era.

\acknowledgments
We thank Ivan Agullo and Kin-Wang Ng for helpful discussions.  This work was supported in part by the National
Science Foundation under Grants No.  PHY-1308325, PHY-1505875, and PHY-1912584 to Wake Forest University.

\appendix

\section{Proof that the Universe is Radiation Dominated at Early Times}\label{sec:proof}

Here we formally prove that in the presence of conformally coupled scalar fields, whether massless or not, to leading order the energy density will scale as $\rho \propto a^{-4}$ at early times for almost any scale factor $a(\eta)$ that vanishes at conformal time $\eta_0$ (which may be finite or $-\infty$) and for a large class of physically acceptable homogeneous and isotropic vacuum states.
As can be seen from~\eqref{RhoAlphaBeta}, this will happen if the integral
\be \label{energy-integral} I(\eta) = \int_0^\infty dk\,k^2 \omega_k(\eta) \left|\beta_k(\eta)\right|^2  \ee
has the property $\lim_{\eta \rightarrow \eta_0} I(\eta) = I(\eta_0)$ with $0 < I(\eta_0) < \infty$.
We begin with the following set of conditions that must be satisfied for the proof to hold:
\bes\bea
\int_0^\infty dk\,k^3 |\beta_k(\eta)|^2 &<& \infty \; , \label{Assume1} \\
\int_0^\lambda dk\, k |\beta_k(\eta_0)|^2 &<& \infty \; , \label{Assume2} \\
\lim_{\eta \rightarrow\eta_0} \left[a^2(\eta)\right]^\prime &<& \infty \; , \label{Assume3} \\
\int_{\eta_0}^\eta dx \left|\left[a^2(x)\right]^{\prime\prime}\right| &<& \infty \; . \label{Assume4}
\eea\ees
We expect that the condition~\eqref{Assume1} will be satisfied by any homogeneous and isotropic vacuum state for which the energy density is
finite for times $\eta > \eta_0$.
Condition~\eqref{Assume2} is satisfied so long as there is at most a weak infrared divergence in $\beta_k$.
Condition~\eqref{Assume3} means the initial expansion is not too extreme, and~\eqref{Assume4} will be automatically satisfied for any universe for which~\eqref{Assume3} applies if $\left[a^2(\eta)\right]^{\prime\prime}$ is always positive or always negative.

To prove that $I(\eta) \rightarrow I(\eta_0)$, we will show that for any $\epsilon >0$, there is a finite time $\bar\eta$ such that for $\eta < \bar\eta$, $|I(\eta) - I(\eta_0)| < \epsilon$.  This can be done by dividing the integral into parts using an infrared cutoff $\lambda$ and then writing $I(\eta)$ as
\be I(\eta) = I(\eta_0) + \Delta I_1(\eta) + \Delta I_2(\eta) + \Delta I_3(\eta) \; , \label{IPieces} \ee
where
\bes\bea
\Delta I_1(\eta) &=& \int_0^\lambda dk \, k^2 \left[\omega_k(\eta) \left| \beta_k(\eta) \right|^2 - k \left|\beta_k(\eta_0)\right|^2 \right] \; ,
\label{I1} \\
\Delta I_2(\eta) &=& \int_\lambda^\infty dk \, k^2 \omega_k(\eta) \left[ \left| \beta_k(\eta) \right|^2 -  \left| \beta_k(\eta_0) \right|^2 \right] \; ,
\label{I2} \\
\Delta I_3(\eta) &=& \int_\lambda^\infty dk \, k^2 \left[\omega_k(\eta) - k \right] \left| \beta_k(\eta_0) \right|^2 \; . \label{I3}
\eea\ees

To place bounds on these quantities we begin by finding inequalities that $\alpha_k$, $\beta_k$, and their first derivatives satisfy.
The inequalities can be obtained using the differential equations satisfied by $\alpha_k$ and $\beta_k$.
 By combining \eqref{AlphaBetaDef1} and \eqref{AlphaBetaDef2}, and substituting the results in \eqref{ModeEquation} with $\xi=\frac16$, one finds
\bes\bea
\alpha_k^\prime &=& \frac{\omega_k^\prime}{2\omega_k} \beta_k e^{2i\theta_k} \; , \label{AlphaPrime} \\
\beta_k^\prime &=& \frac{\omega_k^\prime}{2\omega_k} \alpha_k e^{-2i\theta_k} \; . \label{BetaPrime}
\eea \label{AlphaPrimeBetaPrime}\ees

Using these equations it is straightforward to show that
\be \label{SmallK1}
\frac{\left(\left|\alpha_k\right|^2 +\left|\beta_k\right|^2\right)^\prime}{\left|\alpha_k\right|^2 +\left|\beta_k\right|^2}
= \frac{\omega_k^\prime}{\omega_k} \frac{2{\rm Re}
	\left(\alpha_k \beta_k^* e^{-2i\theta_k}\right)}{\left|\alpha_k\right|^2 +\left|\beta_k\right|^2} \le  \frac{\omega_k^\prime}{\omega_k} \frac{2|\alpha_k| \, |\beta_k|}{\left|\alpha_k\right|^2 +\left|\beta_k\right|^2} < \frac{\omega_k^\prime}{\omega_k}\;,  \ee
where the last inequality can be obtained from  $(|\alpha_k|-|\beta_k|)^2>0$.

Integrating the inequality~\eqref{SmallK1} from $\eta_0$ to $\eta$  yields
\be\label{SmallK2}
\left|\alpha_k(\eta)\right|^2 + \left|\beta_k(\eta)\right|^2 \le \frac{\omega_k(\eta)}{k}\left[ \left|\alpha_k(\eta_0)\right|^2
+ \left|\beta_k(\eta_0)\right|^2 \right] \; .
\ee
Multiplying both sides by $\frac12\omega_k(\eta)$ and using \eqref{WronskianAlphaBeta}, gives
\be
 \omega_k(\eta) \left(\frac12+\left|\beta_k(\eta)\right|^2 \right) \le k \left[1 + \frac{m^2 a^2(\eta)}{k^2}\right] \left[\frac12 + \left| \beta_k(\eta_0)\right|^2 \right] \; .
\ee
Thus it is also true that
 \be\label{SmallK3}
 \omega_k(\eta)\left|\beta_k(\eta)\right|^2  < k \left[1 + \frac{m^2 a^2(\eta)}{k^2}\right] \left[\frac12 + \left| \beta_k(\eta_0)\right|^2 \right] \; . \ee
Integrating this over $k$ up to any finite limit $\lambda$ gives
\be\label{SmallK4}
\int_0^\lambda dk\, k^2 \omega_k(\eta)\left| \beta_k(\eta)\right|^2 < \int_0^\lambda dk\, k \left[k^2 + m^2 a^2(\eta)\right]
\left[\frac12 + \left| \beta_k(\eta_0)\right|^2 \right] \; .
\ee
Note that because of the assumption~\eqref{Assume2}, the integral on the right is finite and is a strictly increasing function of $\eta$.  Choosing an arbitrary conformal time $\eta_1$, restricting to times $\eta < \eta_1$, and choosing $\lambda$ to be small enough allows an arbitrarily small upper bound to be placed on the integral.  Choosing that
bound to be $\frac13 \epsilon$ gives
\be\label{SmallK5}0 \le  \int_0^\lambda dk\, k^2 \omega_k(\eta)\left| \beta_k(\eta)\right|^2 < \frac13\epsilon \quad \hbox{for} \quad \eta<\eta_1 \; . \ee
Equation~\eqref{SmallK5} will be true at all early times, including $\eta =\eta_0$, and it is obviously positive, so comparing with \eqref{I1} we see that
the absolute value of the difference between the integral in~\eqref{SmallK5} for $\eta > \eta_0$ and the integral evaluated at $\eta = \eta_0$ must also satisfy the same bound, so
\be\label{I1Limit}
\left|\Delta I_1(\eta) \right| = \left|\int_0^\lambda dk\, k^2 \omega_k(\eta) \left| \beta_k(\eta)\right|^2
- \int_0^\lambda dk\, k^3  \left| \beta_k(\eta_0)\right|^2\right|< \frac13\epsilon \quad \hbox{for} \quad \eta<\eta_1 \; .
\ee

To make progress on $\Delta I_2$, first note that for \eqref{Assume1} to be satisfied, $|\beta_k(\eta_0)|$ must fall faster than $k^{-2}$ at large values of $k$, and it
must not diverge as quickly as $k^{-2}$ for small values of $k$.  It follows that $k^2|\beta_k(\eta_0)|$ must have an upper bound for all values of $k$, which we call $B$, so that
\be \left| \beta_k(\eta_0) \right| < \frac{B}{k^2}  \; . \label{BetaLimit} \ee
Next~\eqref{BetaPrime} can be integrated to yield
\bes\bea
\beta_k(\eta) &=& \beta_k(\eta_0) + \Delta \beta_k(\eta) \; , \label{DeltaBetaDef} \\
\Delta \beta_k(\eta) &=& \frac12 \int_{\eta_0}^\eta dx \frac{\omega_k^\prime(x)}{\omega_k(x)} \alpha_k(x) e^{-2i\theta_k(x)} \; . \label{DeltaBeta}
\eea\ees
If the condition~\eqref{Assume1} is satisfied and if $\beta_k(\eta)$ is a continuous function of $\eta$ for all $k>0$, then, for any $\lambda>0$ and all $k \ge \lambda$, there will be an upper bound on the value of $|\beta_k(\eta)|$ for $\eta_0 \le \eta < \eta_2$, where $\eta_2 > \eta_0$.
This bound may depend on $\eta_2$, but it will not depend on $k$.
Using~\eqref{WronskianAlphaBeta}, this means that there exist positive constants $\beta_{\rm max}$ and $\alpha_{\rm max}$ such that
\begin{subequations}
	\begin{align}
	\begin{split}
		&|\beta_k(\eta)| < \beta_{\rm{max}} \;, \label{BetaMax}
	\end{split} \\
	\begin{split}
		&\left|\alpha_k(\eta)\right| < \alpha_{\rm max}=\sqrt{1+\beta_{\rm max}^2} \; . \label{AlphaMax}
	\end{split}
	\end{align}
\end{subequations}

Equations~\eqref{DeltaBeta} and \eqref{AlphaMax} can be used to place a limit on $\Delta \beta_k$:
\be
\left|\Delta \beta_k(\eta) \right|
\le \frac12 \int_{\eta_0}^\eta dx\left| \frac{\omega_k^\prime(x)}{\omega_k(x)} \right| \left| \alpha_k(x)\right|
\le \frac{\alpha_{\rm max}}2 \ln\left[ \frac{\omega_k(\eta)}{k} \right]
< \frac{\alpha_{\rm max}}4  \left[ \frac{\omega_k(\eta)}{k} - \frac{k}{\omega_k(\eta)}\right] \; ,
\ee
where the fact that $\omega_k$ is an increasing function of $\eta$ has been used along with the identity $\ln(x)<\frac12 (x-x^{-1})$ when $x>1$.  Thus
\be \left|\Delta \beta_k(\eta) \right| <\frac{\alpha_{\rm max} m^2 a^2(\eta)}{4\,k\,\omega_k(\eta)} \; . \ee
Note that $\Delta \beta_k(\eta)$ vanishes in the limit $\eta \to \eta_0$, so $\Delta \beta_k(\eta)$  can be made arbitrarily small by choosing an early enough time $\eta_2$.
Thus for any $\delta>0$ it is possible to find a time $\eta_2$ such that
\be \left|\Delta \beta_k(\eta) \right| < \frac{\delta}{k\omega_k(\eta)} \le \frac{\delta}{k^2}  \quad \hbox{for} \quad k>\lambda \; , \; \eta_0 \le \eta < \eta_2 \; .\label{DeltaBetaLimit1} \ee

To find a bound on $\Delta I_2(\eta)$, it is useful to derive a second bound on $|\Delta \beta_k(\eta)|$.
It is easy to show that
\begin{equation}
	\Delta \beta_k(\eta) = \frac{i}{4} \int_{\eta_0}^\eta dx 	\frac{\omega_k^\prime(x)}{\omega_k^2(x)} \alpha_k(x)  \frac{d}{dx} \,e^{-2i\theta_k(x)} \;.
\end{equation}
Then, integrating by parts and using~\eqref{AlphaPrime} gives
\begin{align}
	\Delta \beta_k(\eta) =&  \frac{i}{4} \left[ \frac{\omega_k^\prime(\eta)}{\omega_k^2(\eta)} \alpha_k(\eta) e^{-2i\theta_k(\eta)} - \frac{\omega_k^\prime(\eta_0)}{k^2} \alpha_k(\eta_0) e^{-2i\theta_k(\eta_0)}\right] \nonumber\\
	&- \frac{i}{4} \int_{\eta_0}^\eta dx \left\{\left[ \frac{\omega_k^{\prime\prime}(x)}{\omega_k^2(x)} - \frac{2\omega_k^{\prime 2}(x)}{\omega_k^3(x)} \right] \alpha_k(x) e^{-2i \theta_k(x)} + \frac{\omega_k^{\prime 2}(x)}{2\omega_k^3(x)} \beta_k(x) \right\} \; . \label{DeltaBetaParts}
\end{align}
Using \eqref{OmegaDef} one finds that
\bea\label{DeltaBetaParts2}
&& \Delta \beta_k(\eta)= \frac{i m^2}{8} \left[ \frac{\left[a^2(\eta)\right]^\prime}{\omega_k^3(\eta)} \alpha_k(\eta) e^{-2i\theta_k(\eta)}
-\frac{\left[a^2(\eta_0)\right]^\prime}{k^3}   \alpha_k(\eta_0) e^{-2i\theta_k(\eta_0)} \right]
\nonumber \\
&&-  \frac{im^2}{8}\int_{\eta_0}^\eta  dx  \left\{  \frac{ \left[a^2(x)\right]^{\prime\prime} }{\omega_k^3(x)}  \alpha_k(x) e^{-2i\theta_k(x)} - \frac{m^2}{4} \frac{\left[a^2(x)\right]^{\prime 2}}{\omega_k^5(x)} \left[ 6 \alpha_k(x) e^{-2i \theta_k(x)}
- \beta_k(x) \right] \right\}.	
\eea
There might be some concern that $e^{-2i\theta_k(\eta_0)}$ is ill defined in the case $\eta_0 = -\infty$, but in this case it is always true that $[a^2(\eta_0)]^\prime = 0$, so this ambiguity is irrelevant.  Note that condition~\eqref{Assume3} must be satisfied for~\eqref{DeltaBetaParts2} to be a well-defined expression.

We now place a limit on $\Delta \beta_k(\eta)$ using \eqref{DeltaBetaParts2}, \eqref{BetaMax}, \eqref{AlphaMax} and the fact that $\omega_k > k > \lambda$:
\bea
\left|\Delta \beta_k(\eta)\right| &<& \frac{m^2}{8k^3} \alpha_{\rm max} \left\{\left[ a^2(\eta) \right]^\prime + \left[ a^2(\eta_0)\right]^\prime +
\int_{\eta_0}^\eta dx \left| [ a^2(x)]^{\prime\prime} \right| \right\} \nonumber \\
&& \qquad + \frac{m^4}{32k^3\lambda^2}\left(6 \alpha_{\rm max} + \beta_{\rm max}\right) \int_{\eta_0}^\eta dx \left[a^2(x)\right]^{\prime\,2}
\; . \label{DeltaBetaParts3}
\eea
Note that the conditions~\eqref{Assume3} and \eqref{Assume4} ensure that the first three terms are finite, and, therefore, there exists a positive constant $C_1$ such that
\be\label{DeltaBetaPieceA}
\frac{m^2}{8} \alpha_{\rm max} \left\{\left[ a^2(\eta) \right]^\prime + \left[ a^2(\eta_0)\right]^\prime +
\int_{\eta_0}^\eta dx \left| [ a^2(x)]^{\prime\prime} \right| \right\} < C_1 \quad \hbox{for} \quad \eta_0 \le \eta < \eta_2  \; .
\ee
Since $[a^2(x)]^\prime$ is finite as $\eta \rightarrow \eta_0$ its square must be integrable over any finite range.  Thus if $\eta_0$ is finite the last term in \eqref{DeltaBetaParts3} is also finite.  If $\eta_0 = -\infty$, first note that it must be true that $\int_{\eta_0}^\eta [a^2(x)]^\prime dx = a^2(\eta)$ is finite, so $[a^2(\eta)]^\prime$ must fall off faster than $|\eta|^{-1}$ as $\eta \rightarrow - \infty$, and, hence, $[a^2(\eta)]^{\prime \,2}$ falls off faster than $|\eta|^{-2}$ and is also integrable.  Therefore in either case the final term in~\eqref{DeltaBetaParts3} is integrable, and there exists some positive constant $C_2$ such that
\be\label{DeltaBetaPieceB}
\frac{m^4}{32\lambda^2}\left(6 \alpha_{\rm max} + \beta_{\rm max}\right) \int_{\eta_0}^\eta dx \left[ a^2(x)\right]^{\prime2} < C_2
\quad \hbox{for} \quad \eta < \eta_2  \; .
\ee
Substituting \eqref{DeltaBetaPieceA} and \eqref{DeltaBetaPieceB} into \eqref{DeltaBetaParts3}, and noting that for $k > \lambda$ and $\eta < \eta_2$,  $\omega_k(\eta)/k < \omega_\lambda(\eta_2)/\lambda$ one finds that
\be\label{DeltaBetaLimit2a} \left| \Delta \beta_k(\eta) \right| < \frac{C_1+C_2}{k^3} \quad \hbox{for} \quad \eta < \eta_2 \; , \; k>\lambda \; ,\ee
and therefore
\bes \bea\left| \Delta \beta_k(\eta) \right| &<& \frac{D}{k^2\omega_k(\eta)} \quad \hbox{for} \quad \eta < \eta_2 \; , \; k>\lambda \;, \label{DeltaBetaLimit2}  \\
 D & = &   \frac{C_1 + C_2}{\lambda} \omega_\lambda (\eta_2) \;, \label{DDef}
\eea \ees
Combining the two limits~\eqref{DeltaBetaLimit1} and \eqref{DeltaBetaLimit2} gives
\be
\label{DeltaBetaLimit} \left| \Delta \beta_k(\eta) \right| < \frac{1}{\omega_k(\eta)} {\rm min}\left( \frac{\delta}{k},\frac{D}{k^2}\right)
\quad \hbox{for} \quad \eta<\eta_2 \; , \; k> \lambda \; .
\ee

It is possible to put a bound on $|\Delta I_2(\eta)|$ in~\eqref{I2} by
choosing $\delta$ to be small enough so that $\delta \lambda < D$, and using
the bounds in~\eqref{BetaLimit} and~\eqref{DeltaBetaLimit} along with the fact that $\omega_k>k$.  The result is:
\bea\label{I2Step}
\left| \Delta I_2(\eta) \right| &=& \left| \int_\lambda^\infty dk\, k^2 \omega_k(\eta) \left\{ 2 {\rm Re}\left[\beta_k(\eta_0)^* \Delta \beta_k(\eta) \right]
+ \left| \Delta \beta_k(\eta) \right|^2 \right\}\right| \nonumber \\
&<& \int_\lambda^\infty dk\, k^2 \left[ \frac{2B}{k^2} \text{min}\left(\frac{\delta}{k},\frac{D}{k^2}\right)
+ \frac1k \text{min}\left(\frac{\delta}{k},\frac{D}{k^2}\right) ^2 \right]  \; .\eea
The last integral can be computed by dividing it into the integrals $\int_\lambda^{D/\delta} dk + \int_{D/\delta}^\infty dk$ with the result that
\be  \left| \Delta I_2(\eta) \right| <  (2B\delta+\delta^2) \ln \left( \frac{D}{\delta \lambda} \right) + 2B\delta + \frac12 \delta^2 \; .
\ee
Then since $\delta$ can be made as small as desired by choosing $\eta_2$ appropriately, we can use the same bound as in~\eqref{I1Limit}
\be\label{I2Limit} \left| \Delta I_2(\eta) \right| < \frac13\epsilon \quad \hbox{for} \quad \eta<\eta_2 \;. \ee

Finally, we can put a limit on $\Delta I_3$, given by \eqref{I3} by using $\omega_k^2 < (k+ \frac{m^2 a^2}{2 k})^2$ which implies that $\omega_k - k < m^2a^2/2k$, together with \eqref{BetaLimit}, so that
\be
\label{I3Step} |\Delta I_3(\eta)| < \int_\lambda^\infty dk\,k^2 \frac{m^2 a^2(\eta)B^2}{2k^5} =  \frac{m^2 B^2}{4\lambda^2} a^2(\eta) \; .
\ee
We can make this small by simply making $a(\eta)$ small, so we have
\be |\Delta I_3(\eta)| < \frac13 \epsilon   \quad \hbox{for} \quad \eta<\eta_3 \; . \label{I3Limit}\ee
If we then define $\bar \eta = {\rm min}(\eta_1,\eta_2,\eta_3)$ and use \eqref{I1Limit}, \eqref{I2Limit} and \eqref{I3Limit} in \eqref{IPieces} we find
\be\label{ILimit} \left| I(\eta) - I(\eta_0) \right| < \epsilon \quad \hbox{for} \quad \eta<\bar \eta \; . \ee
Since this can be achieved for any $\epsilon > 0$, we conclude that
\be\label{IFinal} \lim_{\eta \rightarrow \eta_0} I(\eta) = I(\eta_0)  = \int_0^\infty dk\,k^3  \left|\beta_k(\eta_0)\right|^2 \; .\ee

\bibliography{RadiationDominatedPreInflation}

\begin{thebibliography}{37}%
\makeatletter
\providecommand \@ifxundefined [1]{%
 \@ifx{#1\undefined}
}%
\providecommand \@ifnum [1]{%
 \ifnum #1\expandafter \@firstoftwo
 \else \expandafter \@secondoftwo
 \fi
}%
\providecommand \@ifx [1]{%
 \ifx #1\expandafter \@firstoftwo
 \else \expandafter \@secondoftwo
 \fi
}%
\providecommand \natexlab [1]{#1}%
\providecommand \enquote  [1]{``#1''}%
\providecommand \bibnamefont  [1]{#1}%
\providecommand \bibfnamefont [1]{#1}%
\providecommand \citenamefont [1]{#1}%
\providecommand \href@noop [0]{\@secondoftwo}%
\providecommand \href [0]{\begingroup \@sanitize@url \@href}%
\providecommand \@href[1]{\@@startlink{#1}\@@href}%
\providecommand \@@href[1]{\endgroup#1\@@endlink}%
\providecommand \@sanitize@url [0]{\catcode `\\12\catcode `\$12\catcode
  `\&12\catcode `\#12\catcode `\^12\catcode `\_12\catcode `\%12\relax}%
\providecommand \@@startlink[1]{}%
\providecommand \@@endlink[0]{}%
\providecommand \url  [0]{\begingroup\@sanitize@url \@url }%
\providecommand \@url [1]{\endgroup\@href {#1}{\urlprefix }}%
\providecommand \urlprefix  [0]{URL }%
\providecommand \Eprint [0]{\href }%
\providecommand \doibase [0]{http://dx.doi.org/}%
\providecommand \selectlanguage [0]{\@gobble}%
\providecommand \bibinfo  [0]{\@secondoftwo}%
\providecommand \bibfield  [0]{\@secondoftwo}%
\providecommand \translation [1]{[#1]}%
\providecommand \BibitemOpen [0]{}%
\providecommand \bibitemStop [0]{}%
\providecommand \bibitemNoStop [0]{.\EOS\space}%
\providecommand \EOS [0]{\spacefactor3000\relax}%
\providecommand \BibitemShut  [1]{\csname bibitem#1\endcsname}%
\let\auto@bib@innerbib\@empty
\bibitem [{\citenamefont {Anderson}(1985)}]{anderson_effects_1985}%
  \BibitemOpen
  \bibfield  {author} {\bibinfo {author} {\bibfnamefont {P.~R.}\ \bibnamefont
  {Anderson}},\ }\href {\doibase 10.1103/PhysRevD.32.1302} {\bibfield
  {journal} {\bibinfo  {journal} {Phys. Rev. D}\ }\textbf {\bibinfo {volume}
  {32}},\ \bibinfo {pages} {1302} (\bibinfo {year} {1985})}\BibitemShut
  {NoStop}%
\bibitem [{\citenamefont {{Wald}}(1978)}]{Wald1978}%
  \BibitemOpen
  \bibfield  {author} {\bibinfo {author} {\bibfnamefont {R.~M.}\ \bibnamefont
  {{Wald}}},\ }\href {https://doi.org/10.1016/0003-4916(78)90040-4} {\bibfield
  {journal} {\bibinfo  {journal} {Ann. Phys. (N. Y.)}\ }\textbf {\bibinfo
  {volume} {110}},\ \bibinfo {pages} {472} (\bibinfo {year}
  {1978})}\BibitemShut {NoStop}%
\bibitem [{\citenamefont {Appignani}\ and\ \citenamefont
  {Casadio}(2008)}]{Appignani_2008}%
  \BibitemOpen
  \bibfield  {author} {\bibinfo {author} {\bibfnamefont {C.}~\bibnamefont
  {Appignani}}\ and\ \bibinfo {author} {\bibfnamefont {R.}~\bibnamefont
  {Casadio}},\ }\href {\doibase 10.1088/1475-7516/2008/10/027} {\bibfield
  {journal} {\bibinfo  {journal} {J. Cosmol. Astropart. Phys.}\ }\textbf
  {\bibinfo {volume} {10}},\ \bibinfo {pages} {027} (\bibinfo {year}
  {2008})}\BibitemShut {NoStop}%
\bibitem [{\citenamefont {Anderson}\ \emph {et~al.}(2005)\citenamefont
  {Anderson}, \citenamefont {Molina-Par\'{\i}s},\ and\ \citenamefont
  {Mottola}}]{anderson_short_2005}%
  \BibitemOpen
  \bibfield  {author} {\bibinfo {author} {\bibfnamefont {P.~R.}\ \bibnamefont
  {Anderson}}, \bibinfo {author} {\bibfnamefont {C.}~\bibnamefont
  {Molina-Par\'{\i}s}}, \ and\ \bibinfo {author} {\bibfnamefont
  {E.}~\bibnamefont {Mottola}},\ }\href {\doibase 10.1103/PhysRevD.72.043515}
  {\bibfield  {journal} {\bibinfo  {journal} {Phys. Rev. D}\ }\textbf {\bibinfo
  {volume} {72}},\ \bibinfo {pages} {043515} (\bibinfo {year}
  {2005})}\BibitemShut {NoStop}%
\bibitem [{\citenamefont {{Nachtmann}}(1967)}]{Nacht}%
  \BibitemOpen
  \bibfield  {author} {\bibinfo {author} {\bibfnamefont {O.}~\bibnamefont
  {{Nachtmann}}},\ }\href {\doibase 10.1007/BF01646319} {\bibfield  {journal}
  {\bibinfo  {journal} {Commun. Math. Phys.}\ }\textbf {\bibinfo {volume}
  {6}},\ \bibinfo {pages} {1} (\bibinfo {year} {1967})}\BibitemShut {NoStop}%
\bibitem [{\citenamefont {{Chernikov}}\ and\ \citenamefont
  {{Tagirov}}(1968)}]{CherTag}%
  \BibitemOpen
  \bibfield  {author} {\bibinfo {author} {\bibfnamefont {N.~A.}\ \bibnamefont
  {{Chernikov}}}\ and\ \bibinfo {author} {\bibfnamefont {E.~A.}\ \bibnamefont
  {{Tagirov}}},\ }\href@noop {} {\bibfield  {journal} {\bibinfo  {journal}
  {Ann. Inst. Henri Poincar\'e A}\ }\textbf {\bibinfo {volume} {9}},\ \bibinfo
  {pages} {109} (\bibinfo {year} {1968})}\BibitemShut {NoStop}%
\bibitem [{\citenamefont {{Tagirov}}(1973)}]{Tag}%
  \BibitemOpen
  \bibfield  {author} {\bibinfo {author} {\bibfnamefont {E.~A.}\ \bibnamefont
  {{Tagirov}}},\ }\href {\doibase 10.1016/0003-4916(73)90047-X} {\bibfield
  {journal} {\bibinfo  {journal} {Ann. Phys. (N.Y.)}\ }\textbf {\bibinfo
  {volume} {76}},\ \bibinfo {pages} {561} (\bibinfo {year} {1973})}\BibitemShut
  {NoStop}%
\bibitem [{\citenamefont {{Bunch}}\ and\ \citenamefont
  {{Davies}}(1978)}]{Bunch-Davies}%
  \BibitemOpen
  \bibfield  {author} {\bibinfo {author} {\bibfnamefont {T.~S.}\ \bibnamefont
  {{Bunch}}}\ and\ \bibinfo {author} {\bibfnamefont {P.~C.~W.}\ \bibnamefont
  {{Davies}}},\ }\href {\doibase https://doi.org/10.1098/rspa.1978.0060}
  {\bibfield  {journal} {\bibinfo  {journal} {Proc. R. Soc. A}\ }\textbf
  {\bibinfo {volume} {360}},\ \bibinfo {pages} {117} (\bibinfo {year}
  {1978})}\BibitemShut {NoStop}%
\bibitem [{\citenamefont {Berera}(1995)}]{warm-inflation}%
  \BibitemOpen
  \bibfield  {author} {\bibinfo {author} {\bibfnamefont {A.}~\bibnamefont
  {Berera}},\ }\href {\doibase https://doi.org/10.1103/PhysRevLett.75.3218}
  {\bibfield  {journal} {\bibinfo  {journal} {Phys. Rev. Lett}\ }\textbf
  {\bibinfo {volume} {75}},\ \bibinfo {pages} {3218} (\bibinfo {year}
  {1995})}\BibitemShut {NoStop}%
\bibitem [{\citenamefont {Parker}(1966)}]{parker-thesis}%
  \BibitemOpen
  \bibfield  {author} {\bibinfo {author} {\bibfnamefont {L.}~\bibnamefont
  {Parker}},\ }\href@noop {} {Ph.D. thesis},\ \bibinfo  {school} {Harvard
  University} (\bibinfo {year} {1966})\BibitemShut {NoStop}%
\bibitem [{\citenamefont {Parker}\ and\ \citenamefont {Fulling}(1974)}]{pf}%
  \BibitemOpen
  \bibfield  {author} {\bibinfo {author} {\bibfnamefont {L.}~\bibnamefont
  {Parker}}\ and\ \bibinfo {author} {\bibfnamefont {S.~A.}\ \bibnamefont
  {Fulling}},\ }\href {\doibase 10.1103/PhysRevD.9.341} {\bibfield  {journal}
  {\bibinfo  {journal} {Phys. Rev. D}\ }\textbf {\bibinfo {volume} {9}},\
  \bibinfo {pages} {341} (\bibinfo {year} {1974})}\BibitemShut {NoStop}%
\bibitem [{\citenamefont {{Fulling}}\ and\ \citenamefont
  {{Parker}}(1974)}]{fp}%
  \BibitemOpen
  \bibfield  {author} {\bibinfo {author} {\bibfnamefont {S.~A.}\ \bibnamefont
  {{Fulling}}}\ and\ \bibinfo {author} {\bibfnamefont {L.}~\bibnamefont
  {{Parker}}},\ }\href {\doibase https://doi.org/10.1016/0003-4916(74)90451-5}
  {\bibfield  {journal} {\bibinfo  {journal} {Ann. Phys. (N. Y.)}\ }\textbf
  {\bibinfo {volume} {87}},\ \bibinfo {pages} {176} (\bibinfo {year}
  {1974})}\BibitemShut {NoStop}%
\bibitem [{\citenamefont {Fulling}\ \emph {et~al.}(1974)\citenamefont
  {Fulling}, \citenamefont {Parker},\ and\ \citenamefont {Hu}}]{fph}%
  \BibitemOpen
  \bibfield  {author} {\bibinfo {author} {\bibfnamefont {S.~A.}\ \bibnamefont
  {Fulling}}, \bibinfo {author} {\bibfnamefont {L.}~\bibnamefont {Parker}}, \
  and\ \bibinfo {author} {\bibfnamefont {B.~L.}\ \bibnamefont {Hu}},\ }\href
  {\doibase 10.1103/PhysRevD.10.3905} {\bibfield  {journal} {\bibinfo
  {journal} {Phys. Rev. D}\ }\textbf {\bibinfo {volume} {10}},\ \bibinfo
  {pages} {3905} (\bibinfo {year} {1974})}\BibitemShut {NoStop}%
\bibitem [{\citenamefont {Bunch}(1980)}]{bunch}%
  \BibitemOpen
  \bibfield  {author} {\bibinfo {author} {\bibfnamefont {T.~S.}\ \bibnamefont
  {Bunch}},\ }\href {\doibase 10.1088/0305-4470/13/4/022} {\bibfield  {journal}
  {\bibinfo  {journal} {J. Phys A}\ }\textbf {\bibinfo {volume} {13}},\
  \bibinfo {pages} {1297} (\bibinfo {year} {1980})}\BibitemShut {NoStop}%
\bibitem [{\citenamefont {Hirai}(2003)}]{hirai_initial_2003}%
  \BibitemOpen
  \bibfield  {author} {\bibinfo {author} {\bibfnamefont {S.}~\bibnamefont
  {Hirai}},\ }\href {\doibase 10.1088/0264-9381/20/9/306} {\bibfield  {journal}
  {\bibinfo  {journal} {Classical and Quantum Gravity}\ }\textbf {\bibinfo
  {volume} {20}},\ \bibinfo {pages} {1673} (\bibinfo {year}
  {2003})}\BibitemShut {NoStop}%
\bibitem [{\citenamefont {Cline}\ \emph {et~al.}(2003)\citenamefont {Cline},
  \citenamefont {Crotty},\ and\ \citenamefont {Lesgourgues}}]{cline_does_2003}%
  \BibitemOpen
  \bibfield  {author} {\bibinfo {author} {\bibfnamefont {J.~M.}\ \bibnamefont
  {Cline}}, \bibinfo {author} {\bibfnamefont {P.}~\bibnamefont {Crotty}}, \
  and\ \bibinfo {author} {\bibfnamefont {J.}~\bibnamefont {Lesgourgues}},\
  }\href {\doibase 10.1088/1475-7516/2003/09/010} {\bibfield  {journal}
  {\bibinfo  {journal} {J. Cosmol. Astropart. Phys.}\ }\textbf {\bibinfo
  {volume} {09}},\ \bibinfo {pages} {010} (\bibinfo {year} {2003})}\BibitemShut
  {NoStop}%
\bibitem [{\citenamefont {Hirai}(2005)}]{hirai_initial_2005}%
  \BibitemOpen
  \bibfield  {author} {\bibinfo {author} {\bibfnamefont {S.}~\bibnamefont
  {Hirai}},\ }\href {\doibase 10.1088/0264-9381/22/7/003} {\bibfield  {journal}
  {\bibinfo  {journal} {Classical and Quantum Gravity}\ }\textbf {\bibinfo
  {volume} {22}},\ \bibinfo {pages} {1239} (\bibinfo {year}
  {2005})}\BibitemShut {NoStop}%
\bibitem [{\citenamefont {Hirai}\ and\ \citenamefont
  {Takami}(2006)}]{hirai_effect_2006}%
  \BibitemOpen
  \bibfield  {author} {\bibinfo {author} {\bibfnamefont {S.}~\bibnamefont
  {Hirai}}\ and\ \bibinfo {author} {\bibfnamefont {T.}~\bibnamefont {Takami}},\
  }\href {\doibase 10.1088/0264-9381/23/7/019} {\bibfield  {journal} {\bibinfo
  {journal} {Classical and Quantum Gravity}\ }\textbf {\bibinfo {volume}
  {23}},\ \bibinfo {pages} {2541} (\bibinfo {year} {2006})}\BibitemShut
  {NoStop}%
\bibitem [{\citenamefont {Powell}\ and\ \citenamefont
  {Kinney}(2007)}]{powell_pre-inflationary-2007}%
  \BibitemOpen
  \bibfield  {author} {\bibinfo {author} {\bibfnamefont {B.~A.}\ \bibnamefont
  {Powell}}\ and\ \bibinfo {author} {\bibfnamefont {W.~H.}\ \bibnamefont
  {Kinney}},\ }\href {\doibase 10.1103/PhysRevD.76.063512} {\bibfield
  {journal} {\bibinfo  {journal} {Phys. Rev. D}\ }\textbf {\bibinfo {volume}
  {76}},\ \bibinfo {pages} {063512} (\bibinfo {year} {2007})}\BibitemShut
  {NoStop}%
\bibitem [{\citenamefont {Hirai}\ and\ \citenamefont
  {Takami}(2007)}]{hirai_effect_2007}%
  \BibitemOpen
  \bibfield  {author} {\bibinfo {author} {\bibfnamefont {S.}~\bibnamefont
  {Hirai}}\ and\ \bibinfo {author} {\bibfnamefont {T.}~\bibnamefont {Takami}},\
  }\href@noop {} {\  (\bibinfo {year} {2007})},\ \Eprint
  {http://arxiv.org/abs/0710.2385} {arXiv:0710.2385} \BibitemShut {NoStop}%
\bibitem [{\citenamefont {Wang}\ and\ \citenamefont
  {Ng}(2008)}]{wang_effects_2008}%
  \BibitemOpen
  \bibfield  {author} {\bibinfo {author} {\bibfnamefont {I.-C.}\ \bibnamefont
  {Wang}}\ and\ \bibinfo {author} {\bibfnamefont {K.-W.}\ \bibnamefont {Ng}},\
  }\href {\doibase 10.1103/PhysRevD.77.083501} {\bibfield  {journal} {\bibinfo
  {journal} {Phys. Rev. D}\ }\textbf {\bibinfo {volume} {77}},\ \bibinfo
  {pages} {083501} (\bibinfo {year} {2008})}\BibitemShut {NoStop}%
\bibitem [{\citenamefont {{Nicholson}}\ and\ \citenamefont
  {{Contaldi}}(2008)}]{Nicholson2008_The}%
  \BibitemOpen
  \bibfield  {author} {\bibinfo {author} {\bibfnamefont {G.}~\bibnamefont
  {{Nicholson}}}\ and\ \bibinfo {author} {\bibfnamefont {C.~R.}\ \bibnamefont
  {{Contaldi}}},\ }\href {\doibase 10.1088/1475-7516/2008/01/002} {\bibfield
  {journal} {\bibinfo  {journal} {J. Cosmol. Astropart. Phys.}\ }\textbf
  {\bibinfo {volume} {01}},\ \bibinfo {pages} {002} (\bibinfo {year} {2008})},\
  \Eprint {http://arxiv.org/abs/astro-ph/0701783} {arXiv:astro-ph/0701783}
  \BibitemShut {NoStop}%
\bibitem [{\citenamefont {Marozzi}\ \emph {et~al.}(2011)\citenamefont
  {Marozzi}, \citenamefont {Rinaldi},\ and\ \citenamefont
  {Durrer}}]{marozzi_on_2011}%
  \BibitemOpen
  \bibfield  {author} {\bibinfo {author} {\bibfnamefont {G.}~\bibnamefont
  {Marozzi}}, \bibinfo {author} {\bibfnamefont {M.}~\bibnamefont {Rinaldi}}, \
  and\ \bibinfo {author} {\bibfnamefont {R.}~\bibnamefont {Durrer}},\ }\href
  {\doibase 10.1103/PhysRevD.83.105017} {\bibfield  {journal} {\bibinfo
  {journal} {Phys. Rev. D}\ }\textbf {\bibinfo {volume} {83}},\ \bibinfo
  {pages} {105017} (\bibinfo {year} {2011})}\BibitemShut {NoStop}%
\bibitem [{\citenamefont {Cicoli}\ \emph {et~al.}(2014)\citenamefont {Cicoli},
  \citenamefont {Downes}, \citenamefont {Dutta}, \citenamefont {Pedro},\ and\
  \citenamefont {Westphal}}]{cicoli_just_2014}%
  \BibitemOpen
  \bibfield  {author} {\bibinfo {author} {\bibfnamefont {M.}~\bibnamefont
  {Cicoli}}, \bibinfo {author} {\bibfnamefont {S.}~\bibnamefont {Downes}},
  \bibinfo {author} {\bibfnamefont {B.}~\bibnamefont {Dutta}}, \bibinfo
  {author} {\bibfnamefont {F.~G.}\ \bibnamefont {Pedro}}, \ and\ \bibinfo
  {author} {\bibfnamefont {A.}~\bibnamefont {Westphal}},\ }\href {\doibase
  10.1088/1475-7516/2014/12/030} {\bibfield  {journal} {\bibinfo  {journal} {J.
  Cosmol. Astropart. Phys.}\ }\textbf {\bibinfo {volume} {14}},\ \bibinfo
  {pages} {030} (\bibinfo {year} {2014})}\BibitemShut {NoStop}%
\bibitem [{\citenamefont {Das}\ \emph {et~al.}(2015)\citenamefont {Das},
  \citenamefont {Goswami}, \citenamefont {Prasad},\ and\ \citenamefont
  {Rangarajan}}]{das_revisiting_2015}%
  \BibitemOpen
  \bibfield  {author} {\bibinfo {author} {\bibfnamefont {S.}~\bibnamefont
  {Das}}, \bibinfo {author} {\bibfnamefont {G.}~\bibnamefont {Goswami}},
  \bibinfo {author} {\bibfnamefont {J.}~\bibnamefont {Prasad}}, \ and\ \bibinfo
  {author} {\bibfnamefont {R.}~\bibnamefont {Rangarajan}},\ }\href {\doibase
  10.1088/1475-7516/2015/06/001} {\bibfield  {journal} {\bibinfo  {journal} {J.
  Cosmol. Astropart. Phys.}\ }\textbf {\bibinfo {volume} {06}},\ \bibinfo
  {pages} {001} (\bibinfo {year} {2015})}\BibitemShut {NoStop}%
\bibitem [{\citenamefont {Das}\ \emph {et~al.}(2016)\citenamefont {Das},
  \citenamefont {Goswami}, \citenamefont {Prasad},\ and\ \citenamefont
  {Rangarajan}}]{Das2016_Constraints}%
  \BibitemOpen
  \bibfield  {author} {\bibinfo {author} {\bibfnamefont {S.}~\bibnamefont
  {Das}}, \bibinfo {author} {\bibfnamefont {G.}~\bibnamefont {Goswami}},
  \bibinfo {author} {\bibfnamefont {J.}~\bibnamefont {Prasad}}, \ and\ \bibinfo
  {author} {\bibfnamefont {R.}~\bibnamefont {Rangarajan}},\ }\href {\doibase
  10.1103/PhysRevD.93.023516} {\bibfield  {journal} {\bibinfo  {journal} {Phys.
  Rev. D}\ }\textbf {\bibinfo {volume} {93}},\ \bibinfo {pages} {023516}
  (\bibinfo {year} {2016})}\BibitemShut {NoStop}%
\bibitem [{\citenamefont {Birrell}\ and\ \citenamefont
  {Davies}(1982)}]{birrell_quantum_1982}%
  \BibitemOpen
  \bibfield  {author} {\bibinfo {author} {\bibfnamefont {N.~D.}\ \bibnamefont
  {Birrell}}\ and\ \bibinfo {author} {\bibfnamefont {P.~C.~W.}\ \bibnamefont
  {Davies}},\ }\href {\doibase 10.1017/CBO9780511622632} {\emph {\bibinfo
  {title} {Quantum Fields in Curved Space}}},\ Cambridge Monographs on
  Mathematical Physics\ (\bibinfo  {publisher} {Cambridge University Press,
  Cambridge, England},\ \bibinfo {year} {1982})\BibitemShut {NoStop}%
\bibitem [{\citenamefont {Starobinsky}(1980)}]{starobinsky_new_1980}%
  \BibitemOpen
  \bibfield  {author} {\bibinfo {author} {\bibfnamefont {A.~A.}\ \bibnamefont
  {Starobinsky}},\ }\href {\doibase 10.1016/0370-2693(80)90670-X} {\bibfield
  {journal} {\bibinfo  {journal} {Phys. Lett. B}\ }\textbf {\bibinfo {volume}
  {91}},\ \bibinfo {pages} {99} (\bibinfo {year} {1980})}\BibitemShut {NoStop}%
\bibitem [{\citenamefont {Anderson}\ and\ \citenamefont
  {Eaker}(1999)}]{Anderson_1999}%
  \BibitemOpen
  \bibfield  {author} {\bibinfo {author} {\bibfnamefont {P.~R.}\ \bibnamefont
  {Anderson}}\ and\ \bibinfo {author} {\bibfnamefont {W.}~\bibnamefont
  {Eaker}},\ }\href {http://dx.doi.org/10.1103/PhysRevD.61.024003} {\bibfield
  {journal} {\bibinfo  {journal} {Phys. Rev. D}\ }\textbf {\bibinfo {volume}
  {61}},\ \bibinfo {pages} {024003} (\bibinfo {year} {1999})}\BibitemShut
  {NoStop}%
\bibitem [{\citenamefont {Anderson}\ and\ \citenamefont
  {Parker}(1987)}]{Anderson1987_Adiabatic}%
  \BibitemOpen
  \bibfield  {author} {\bibinfo {author} {\bibfnamefont {P.~R.}\ \bibnamefont
  {Anderson}}\ and\ \bibinfo {author} {\bibfnamefont {L.}~\bibnamefont
  {Parker}},\ }\href {\doibase 10.1103/PhysRevD.36.2963} {\bibfield  {journal}
  {\bibinfo  {journal} {Phys. Rev. D}\ }\textbf {\bibinfo {volume} {36}},\
  \bibinfo {pages} {2963} (\bibinfo {year} {1987})}\BibitemShut {NoStop}%
\bibitem [{\citenamefont {Agullo}\ \emph {et~al.}(2015)\citenamefont {Agullo},
  \citenamefont {Nelson},\ and\ \citenamefont
  {Ashtekar}}]{Agullo-Nelson-Ashtekar}%
  \BibitemOpen
  \bibfield  {author} {\bibinfo {author} {\bibfnamefont {I.}~\bibnamefont
  {Agullo}}, \bibinfo {author} {\bibfnamefont {W.}~\bibnamefont {Nelson}}, \
  and\ \bibinfo {author} {\bibfnamefont {A.}~\bibnamefont {Ashtekar}},\ }\href
  {\doibase 10.1103/PhysRevD.91.064051} {\bibfield  {journal} {\bibinfo
  {journal} {Phys. Rev. D}\ }\textbf {\bibinfo {volume} {91}},\ \bibinfo
  {pages} {064051} (\bibinfo {year} {2015})}\BibitemShut {NoStop}%
\bibitem [{\citenamefont {{Abramowitz}}\ and\ \citenamefont
  {{Stegun}}(1964)}]{abramowitz_stegun_handbook_1964}%
  \BibitemOpen
  \bibfield  {author} {\bibinfo {author} {\bibfnamefont {M.}~\bibnamefont
  {{Abramowitz}}}\ and\ \bibinfo {author} {\bibfnamefont {I.~A.}\ \bibnamefont
  {{Stegun}}},\ }\href@noop {} {\emph {\bibinfo {title} {Handbook of
  Mathematical Functions with Formulas, Graphs, and Mathematical Tables}}}\
  (\bibinfo  {publisher} {Dover},\ \bibinfo {address} {New York},\ \bibinfo
  {year} {1964})\BibitemShut {NoStop}%
\bibitem [{\citenamefont {{Riotto}}()}]{Riotto2002_Inflation}%
  \BibitemOpen
  \bibfield  {author} {\bibinfo {author} {\bibfnamefont {A.}~\bibnamefont
  {{Riotto}}},\ }\href@noop {} {}\Eprint {http://arxiv.org/abs/hep-ph/0210162}
  {arXiv:hep-ph/0210162} \BibitemShut {NoStop}%
\bibitem [{\citenamefont {{Bennett}}\ \emph {et~al.}(2003)\citenamefont
  {{Bennett}}, \citenamefont {{Halpern}}, \citenamefont {{Hinshaw}},
  \citenamefont {{Jarosik}}, \citenamefont {{Kogut}}, \citenamefont {{Limon}},
  \citenamefont {{Meyer}}, \citenamefont {{Page}}, \citenamefont {{Spergel}},
  \citenamefont {{Tucker}}, \citenamefont {{Wollack}}, \citenamefont
  {{Wright}}, \citenamefont {{Barnes}}, \citenamefont {{Greason}},
  \citenamefont {{Hill}}, \citenamefont {{Komatsu}}, \citenamefont {{Nolta}},
  \citenamefont {{Odegard}}, \citenamefont {{Peiris}}, \citenamefont
  {{Verde}},\ and\ \citenamefont {{Weiland}}}]{Bennet2003_First}%
  \BibitemOpen
  \bibfield  {author} {\bibinfo {author} {\bibfnamefont {C.~L.}\ \bibnamefont
  {{Bennett}}}, \bibinfo {author} {\bibfnamefont {M.}~\bibnamefont
  {{Halpern}}}, \bibinfo {author} {\bibfnamefont {G.}~\bibnamefont
  {{Hinshaw}}}, \bibinfo {author} {\bibfnamefont {N.}~\bibnamefont
  {{Jarosik}}}, \bibinfo {author} {\bibfnamefont {A.}~\bibnamefont {{Kogut}}},
  \bibinfo {author} {\bibfnamefont {M.}~\bibnamefont {{Limon}}}, \bibinfo
  {author} {\bibfnamefont {S.~S.}\ \bibnamefont {{Meyer}}}, \bibinfo {author}
  {\bibfnamefont {L.}~\bibnamefont {{Page}}}, \bibinfo {author} {\bibfnamefont
  {D.~N.}\ \bibnamefont {{Spergel}}}, \bibinfo {author} {\bibfnamefont {G.~S.}\
  \bibnamefont {{Tucker}}}, \bibinfo {author} {\bibfnamefont {E.}~\bibnamefont
  {{Wollack}}}, \bibinfo {author} {\bibfnamefont {E.~L.}\ \bibnamefont
  {{Wright}}}, \bibinfo {author} {\bibfnamefont {C.}~\bibnamefont {{Barnes}}},
  \bibinfo {author} {\bibfnamefont {M.~R.}\ \bibnamefont {{Greason}}}, \bibinfo
  {author} {\bibfnamefont {R.~S.}\ \bibnamefont {{Hill}}}, \bibinfo {author}
  {\bibfnamefont {E.}~\bibnamefont {{Komatsu}}}, \bibinfo {author}
  {\bibfnamefont {M.~R.}\ \bibnamefont {{Nolta}}}, \bibinfo {author}
  {\bibfnamefont {N.}~\bibnamefont {{Odegard}}}, \bibinfo {author}
  {\bibfnamefont {H.~V.}\ \bibnamefont {{Peiris}}}, \bibinfo {author}
  {\bibfnamefont {L.}~\bibnamefont {{Verde}}}, \ and\ \bibinfo {author}
  {\bibfnamefont {J.~L.}\ \bibnamefont {{Weiland}}},\ }\href {\doibase
  10.1086/377253} {\bibfield  {journal} {\bibinfo  {journal} {Astrophys. J.
  Suppl. Ser.}\ }\textbf {\bibinfo {volume} {148}},\ \bibinfo {pages} {1}
  (\bibinfo {year} {2003})}\BibitemShut {NoStop}%
\bibitem [{\citenamefont {Bedroya}\ and\ \citenamefont
  {Vafa}()}]{bedroya2019transplanckian}%
  \BibitemOpen
  \bibfield  {author} {\bibinfo {author} {\bibfnamefont {A.}~\bibnamefont
  {Bedroya}}\ and\ \bibinfo {author} {\bibfnamefont {C.}~\bibnamefont {Vafa}},\
  }\href@noop {} {}\Eprint {http://arxiv.org/abs/1909.11063} {arXiv:1909.11063}
  \BibitemShut {NoStop}%
\bibitem [{\citenamefont {Bedroya}\ \emph {et~al.}(2020)\citenamefont
  {Bedroya}, \citenamefont {Brandenberger}, \citenamefont {Loverde},\ and\
  \citenamefont {Vafa}}]{Bedroya_2020}%
  \BibitemOpen
  \bibfield  {author} {\bibinfo {author} {\bibfnamefont {A.}~\bibnamefont
  {Bedroya}}, \bibinfo {author} {\bibfnamefont {R.}~\bibnamefont
  {Brandenberger}}, \bibinfo {author} {\bibfnamefont {M.}~\bibnamefont
  {Loverde}}, \ and\ \bibinfo {author} {\bibfnamefont {C.}~\bibnamefont
  {Vafa}},\ }\href {\doibase 10.1103/physrevd.101.103502} {\bibfield  {journal}
  {\bibinfo  {journal} {Phys. Rev. D}\ }\textbf {\bibinfo {volume} {101}},\
  \bibinfo {pages} {103502} (\bibinfo {year} {2020})}\BibitemShut {NoStop}%
\bibitem [{\citenamefont {Berera}\ \emph {et~al.}(2020)\citenamefont {Berera},
  \citenamefont {Brahma},\ and\ \citenamefont {Calderón}}]{berera2020role}%
  \BibitemOpen
  \bibfield  {author} {\bibinfo {author} {\bibfnamefont {A.}~\bibnamefont
  {Berera}}, \bibinfo {author} {\bibfnamefont {S.}~\bibnamefont {Brahma}}, \
  and\ \bibinfo {author} {\bibfnamefont {J.~R.}\ \bibnamefont {Calderón}},\
  }\href@noop {} {\bibfield  {journal} {\bibinfo  {journal} {J. High Energy
  Phys.}\ }\textbf {\bibinfo {volume} {08}},\ \bibinfo {pages} {071} (\bibinfo
  {year} {2020})},\ \Eprint {http://arxiv.org/abs/2003.07184}
  {arXiv:2003.07184} \BibitemShut {NoStop}%
\end{thebibliography}%
\end{document}